\documentclass[12pt]{article}
\usepackage{amsmath}
\usepackage{graphicx,psfrag,epsf}
\usepackage{enumerate}
\usepackage{rotating} 
\usepackage{color}
\usepackage{mathtools}

\usepackage{graphicx, multirow}
\usepackage{amsmath,amssymb}

\usepackage{mathtools}
\usepackage{nicefrac}
\usepackage{appendix}
\usepackage{url}
\usepackage{bm}  
\usepackage[round]{natbib}

\graphicspath{{../figs}{./}}

\definecolor{Dblue}{rgb}{0.1,0,0.55}
\definecolor{mygreen}{rgb}{0.1,0.6,0.2}
\definecolor{myorange}{rgb}{0.96, 0.49, 0.26}
\definecolor{turquoise}{rgb}{0.19, 0.84, 0.78}

\newcommand{\red}[1]{\textcolor{black}{#1}}

\newcommand{\blue}[1]{\textcolor{black}{#1}}

\newcommand{\sR}{\hbox{I\kern-.1667em\hbox{R}}}

\usepackage[margin=1in]{geometry}
\usepackage{authblk}

\def\b0{\mbox{\bf{0}}}

\usepackage{float}
\usepackage{setspace}
\DeclareMathOperator*{\argmin}{argmin}
\DeclareMathOperator*{\argmax}{argmax}
\renewcommand\l{\left}
\renewcommand\r{\right}

\newcommand{\bt}[1]{\bm{#1}}

\graphicspath{{./figs/}{../figs/}}

\begin{document}


\singlespacing
\title{\textbf{Sequential Design of Mixture Experiments with an
Empirically  Determined Input Domain and an
Application to Burn-up Credit  Penalization of Nuclear Fuel Rods
}}
\author[1]{Fran\c{c}ois Bachoc\footnote{Corresponding author. Institut de Math\'ematiques de Toulouse, 118 route de Narbonne, 31062 Toulouse, France. francois.bachoc@math.univ-toulouse.fr}}
\author[2]{Th\'{e}o Barthe}
\author[3]{Thomas Santner}
\author[4]{Yann Richet}
\affil[1]{Institut de Math\'{e}matiques de Toulouse, 118 route de Narbonne, 31062, Toulouse, France.}
\affil[2]{Atos, Les Espaces St Martin, 6 Impasse Alice Guy, 31300 Toulouse, France}
\affil[3]{The Ohio State University, 1958 Neil Avenue,
Columbus, Ohio 43210, United States}
\affil[4]{Institut de Radioprotection et de S\^{u}ret\'{e}
        Nucl\'{e}aire, 31 Avenue de la Division Leclerc, 92260 Fontenay-aux-Roses, France}


\date{}

\date{}
\maketitle


\doublespacing 


\begin{abstract}

This paper proposes a sequential design for maximizing a 
stochastic computer simulator
output, $y(\bm{x})$, over an {\em unknown optimization domain}.    The training data
used to estimate the  optimization domain 
are a set of (historical) inputs, often from a physical system
modeled by the simulator.  Two methods
are provided for estimating the simulator input
domain.  An extension of the
well-known efficient global optimization
algorithm is presented to maximize $y(\bm{x})$.  The 
domain estimation/maximization procedure is applied to 
two readily understood analytic examples. It is also used to solve
 a problem in nuclear safety by maximizing the k-effective
``criticality coefficient" of spent fuel rods, considered as one-dimensional heterogeneous fissile media. 
One of the two domain estimation methods relies on expertise-type constraints. We show that these constraints, 
initially chosen to
 address the spent fuel rod example, are robust in that they also lead to good results 
in the second analytic optimization example. 
Of course, in other applications, it could be necessary to design 
alternative constraints that are more suitable for these applications. 

\end{abstract}





\section{Introduction}
\label{sec:intro}

Among the important issues in safety assessment is the prevention
of accidental events. In nuclear safety applications, there are at least two  ways 
of minimizing potential 
accidental events:  the identification of 
worst cases  (and then averting of such cases), and the 
probabilistic containment of accident consequences.  Depending on the  application, the opportunity to 
choose one or the other method can rely on practical considerations, but should be a
consistent part of the whole safety framework and include  information from 
many fields (say, for example, seismology, structural mechanics, nuclear core cooling, neutronics, and radiology).

Common industrial applications have fewer uncontrollable external conditions 
than applications subject
to environmental factors. 
Indeed, sophisticated mathematical models of industrial safety studies are ordinarily regarded as reliable 
descriptions of their performance in the real-world.  Thus, 
many industrial safety studies use mathematical models of the industrial
process to  identify and then avoid worst-case scenarios. 

However when the complexity of the safety study increases and the mathematical model is sensitive to 
uncertain parameters, the prevention of mathematically-determined 
unacceptable events becomes a less reliable method of preventing accidental events. 
A typical example of increasing system complexity occurs when the 
known homogeneity of a critical materials' density, 
its mixing phases, its temperature, or  other spatially-dependent properties can 
not be guaranteed to be  assumed fixed values.
To more accurately approximate reality, the homogeneous model of critical 
components must be replaced by an 
imperfectly-known, heterogeneous one. However, 
it is typically far more difficult to determine the worst case 
performance of a system having heterogeneous components than 
\blue{systems having known subsystems}.

This paper proposes methodology to provide a worst-but-credible-case 
for imperfectly known heterogeneous  models.  The methodology
is illustrated in analytical examples and in an application to
nuclear fuel storage. In the latter example,
an assessment is made of the stability of fissile fuel rods 
after their previous use
in a nuclear reactor   (their ``burn-up credit'')
in order  to relax their 
storage requirements.  Fissile fuel rods identified as 
more stable can be stored in a reduced-risk facility.

The goals of this paper are two-fold.
First, it estimates the (optimization) domain 
${\cal X}$ of inputs ``consistent'' with a training set of 
inputs, say $\bm{x}_1, \ldots, \bm{x}_n$, possibly historical data from 
a physical system with the same domain
as the simulator.  Second, it identifies an $\bm{x}^{\star} \in {\cal X}$
that maximizes $y(\bm{x})$
over $\bm{x} \in {\cal X}$.


The literature contains a number of papers that 
provide additional relevant background 
useful to more fully understand the nuclear safety example
and the statistical optimization of stochastic
simulators.
%
\citet{cacuci2010handbook} provides basic grounding on nuclear engineering and on 
the numerical simulation for such applications. 
\citet{StiHerSte2003} and \citet{DraSanDea2012} propose
statistical methodology for constructing input designs
for simulators that have bounded polygonal
input domains.  \citet{Kle2008} 
reviews the optimization of a deterministic function
defined on a simplex.
\blue{\citet{PicGinRic2013}} estimate sequentially the quantile of
a function $y(\bf{X})$ with random inputs $\bm{X}$ when $y(\bf{x})$ 
is observed with measurement error.


The remainder of this paper is organized as follows.
Section~\ref{sec:MathDesc} states the mathematical notation
used to formally describe the problems solved in subsequent
sections of the paper. 
Section~\ref{sec:sequentialoptimization} reviews the Efficient
Global Optimization (EGO) of \citet{JonSchWel1998} for
minimizing an unknown
$y( \bm{x})$
$:\mathcal{X} \mapsto \mathbb{R}$ over a
rectangular $\mathcal{X}$ and modifications
of  EGO for cases when $y(\bm{x})$ is measured with noise.
Section~\ref{sec:computation} introduces
two methods for identifying a set of inputs
$\bm{x}$ that are
compatible with the training inputs.  One method,
given in Subsection~\ref{subsec:expert},
uses expert-type constraints and a second method,
described in Subsection~\ref{subsec:kernel},
uses a
kernel density estimation approach.
 Finally, Section~\ref{sec:examples}
gives three examples; the first is an easily understood analytic
application which is used to observe the performance of
the proposed methodology; the second is a determination of
configurations of spent fuel rods in nuclear power
reactors that are associated with high criticality settings and the third is an analytical example that illustrates the robustness and generalizability of the global methodology.


\section{Mathematical Description of the Optimization Problem}
\label{sec:MathDesc}

First, the mathematical notation and assumptions will be stated and
then the nuclear safety  application will be stated to illustrate the notation.
 \blue{Consider}  a real-valued (simulator) $y(\bt{\cdot})$
having functional input $x(t)$;   
$x(t)$ is assumed to be {\em positive} and {\em continuous}
with argument $t$ having domain  that is
a bounded interval  that is  taken to be
$[0,1]$, possibly after a location shift and scale transformation. 
Let  $\bm{x}$ denote the input function $ \{ x(t) \}_{t \in [0,1]}$.  
In our nuclear safety application, all  inputs  $\bm{x}$   
are assumed to come from a domain determined by
a  training 
set of inputs to a physical system that is described
below in more detail. 
The simulator output at input $\bm{x}$ is 
$y(\bm{x})$ corrupted by an additive measurement error.

Two other features of the input functions that are consistent with 
nuclear safety study will be assumed.  First, all  inputs $x(t)$ 
are measured  at a common finite grid of $d$ values, say,
$0 \le t_1 < \cdots < t_d \le 1$. This grid
is assumed to be
the {\em same} for all functions.  A linear, quadratic or other 
interpolation scheme would be applied to the available 
$x(t)$ measurements to achieve  a common $t$ grid if this is not true on their native scale. 
Because of their origin, this paper will  refer to the 
$d \times 1$ vector $\bm{x} = \left(x(t_1),\ldots,  x(t_d) \right)^\top$
as a curve or a function. 
Second, by dividing each component of
$\bm{x} $
by its mean $\bar{x} =\frac{1}{d}
\sum_{j=1}^{d}x(t_j)$  it is assumed that $\sum_{j=1}^{d} x(t_j) = d$ for all inputs. 
Thus the set of valid inputs is a {\em subset} of the 
positive $d$-hyperplane, i.e., of 
$\left\{ (w_1,\ldots,w_d): w_j \geq 0 \mbox{\ for } j = 1,\dots,d; \sum_{j=1}^d w_j = d \right\}$. 
 Equivalently, the $d \times 1$ vector $\bm{x}$
has non-negative discrete values whose \blue{\em average} is one.

\begin{figure}[ht]
  \begin{center}
  \includegraphics[width=8cm]{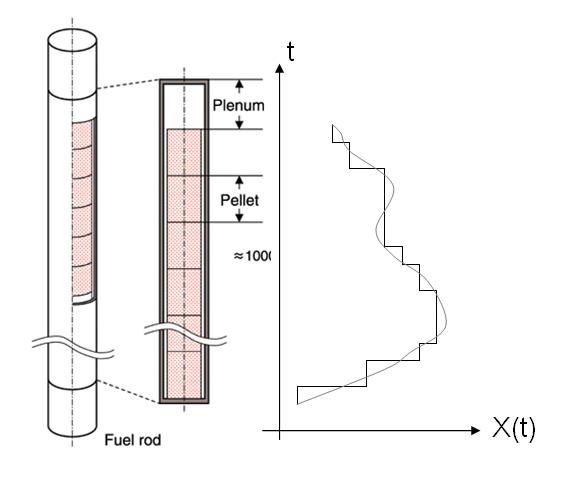}
  \end{center}
\caption{Schematic of a spent fuel rod and its
burn-up rate curve.}
  \label{fig:spentfuelrod}
  \end{figure}

To illustrate the notation of the previous two paragraphs 
for the nuclear safety application, consider 
spent fuel rods that are \red{retrieved} from a nuclear reactor
and inspected. These rods are modeled as one-dimensional heterogeneous fissile media.
Here  $x(t)$ is the ``burn-up rate''
(in \red{megawatt-days/ton}) measured
at vertical position $t$ along the fuel rod,
where the rod is scaled so
that $t \in [0,1]$. 
Remark that the fuel rods are considered for storage at a fixed time, thus the burn-up rate is considered to depend only on the position and not on the time. 
Figure~\ref{fig:spentfuelrod}
is a cartoon that illustrates a spent fuel rod
and the corresponding burn-up rate
energy $x(t)$.
In this application we have available $n =$ 3,158
spent fuel rods (and their burn-up rate curves) from standard nuclear power plants,
(\citet{pwrbup1997}).  The
burn-up rate is measured at $d = 18$
equally-spaced points along $[0,1]$.
Let $\bm{x}_i = \left( x_i(t_1),\ldots,x_i(t_{18})\right)^\top$, $i = 1,\ldots,$ 3,158 denote 
the burn-up rate curves for the  spent fuel rods. 
Each $\bm{x}_i$ 
is an element of the positive $18$-hyperplane.

For any input  $\bm{x}$,
the associated simulator output $y(\bm{x})$ is 
the {\em criticality coefficient} at   $\bm{x}$.  
The {\em criticality coefficient} is computed from 
 depletion calculations  made for each of the $d = 18$ zones  of the rod
 based on macroscopic cross-sections.  The  zone-specific  determinations
 are made
using the 
numerical simulation package {\tt CRISTAL} 
\citep{CRISTAL} and is
followed by a Monte-Carlo $k$-effective calculation for the 
entire burn-up rate curve. 
The $y(\bm{x})$ value is interpreted as follows: 
if $y(\bm{x})< 1$, the rod is called ``subcritical'';  if $y(\bm{x}) = 1$, the rod
is ``critical''; and if $y(\bm{x})> 1$, the rod is
termed ``super-critical''. In particular, an higher value of $y(\bm{x})$ corresponds to more risk and thus the goal is to maximize $y(\bm{x})$, for safety study.

Unfortunately, the observed value of the criticality coefficient
is a noisy version of $y(\bm{x})$. In addition,
{\tt CRISTAL} 
can be costly to run although, 
\red{in this application}, the evaluations
are approximately 15 minutes each.
Let $y_i = y(\bm{x}_i)$ denote the observed 
computed criticality coefficient
for the $i^{th}$  spent fuel rod,
$i = 1,\ldots,$ 3,158.


\section{Sequential Optimization} 
\label{sec:sequentialoptimization}


Section~\ref{sec:computation} will describe two methods for
constructing an input domain ${\cal X}$ that is consistent with historical input data to
the desired  physical system. 
This section will review the ``expected 
improvement'' sequential designs of
\cite{SchWelJon1998} and \cite{JonSchWel1998}
that were introduced to minimize a deterministic $y(\bm{x})$ 
when evaluations are costly. 
Called {\em Efficient Global Optimization} (EGO)
algorithms these designs seek to find
an $x_{\min} \in \arg\min_{\bm{x} \in \mathcal{X}} y(\bm{x})$.
The problem  of maximizing $y(\bm{x})$
 can be solved by applying EGO 
to minimize $- y(\bm{x})$. 
Additional modifications will be given to handle  cases when $y(\bm{x})$ 
is measured with error. 

In brief, EGO is initiated by computing $y(\bm{x})$ 
on a space-filling set of inputs of ${\cal X}$.  
Thus initial information about $y(\bm{\cdot})$ is available 
over a wide, if not dense, subset of the input space.  

At each  update step, EGO  
adds one input $\bm{x} \in {\cal X}$ to the previous
design and the associated $y(\bm{x})$ to the output vector.
Suppose that there have been 
previous evaluations at $y(\bm{x}_1)$, \ldots, $y(\bm{x}_n)$. 
EGO identifies the  next input at which to evaluate
$y(\bm{\cdot})$, denoted $\bm{x}_{n+1}$, 
as the  $\bm{x}$ which maximizes the {\em idealized 
improvement function}
\begin{equation}
\label{eq6.3.theorimp}
{\cal I}(\bm{x})
= 
\left\{
\begin{array}{ll}
y_{\min}^n - y(\bm{x}), & y_{\min}^n - y(\bm{x}) > 0  \\[1ex]
0, & y_{\min}^n - y(\bm{x}) \leq 0
\end{array} 
\right. 
\end{equation}
where 
$y_{\min}^n = \min_{i=1,\ldots,n} y(\bm{x}_i)$
is the smallest value of $y(\bm{x})$ among the  
previous evaluations.  Intuitively, larger values of ${\cal I}(\bm{x})$
produce smaller values of $y(\bm{x})$.  

While $y_{\min}^n$ 
is known, both
$y(\bm{x})$   and hence ${\cal I}(\bm{x})$ are  unknown. 
 EGO uses a Gaussian process extension of 
the regression predictor to estimate $y(\bm{x})$ by $\widehat{y}(\bm{x})$ say, 
and to quantify the uncertainty in this predictor by $s(\bm{x})$, say
(See \citet{SchWelJon1998} 
or Chapter~3 of \citet{SanWilNot2018}).  This stochastic approximation
can be used to find a formula for the expected value  of a stochastic
version of ${\cal I}(\bm{x})$ given the current data. 
The resulting (practical) improvement 
function is
\begin{align}
\MoveEqLeft EI\l[ (\bm{x})  \r]
 = 
(y_{\min}^n - \widehat{y}(\bm{x})) \, 
\Phi \left( \frac{y_{\min}^n - \widehat{y}(\bm{x})}{s(\bm{x})} \right)+ 
s(\bm{x}) \, \phi \left(
\frac{y_{\min}^n - \widehat{y}(\bm{x})}{s(\bm{x})} \right),
\label{eqEXPIMP}
\end{align}
where $\Phi (\bm{\cdot})$ and $\phi (\bm{\cdot})$ are the $N(0,1)$
distribution and density function, respectively.

\smallskip

\blue{EGO is typically
stopped after a fixed budget has been exhausted for 
$y(\bm{x})$ evaluations or 
when the maximum expected improvement is ``small''. 
When EGO stops sampling, it
predicts $\bm{x}_{\min}$ to be that member of}
the current
set of inputs at which $y(\bf{\cdot})$ has been evaluated, say 
$\{\bm{x}_1,\ldots,\bm{x}_{N} \}$, 
to satisfy
\begin{equation}
\label{eq6.3.2001}
y(\widehat{\bm{x}}_{\min})
=
\min_{i=1,\ldots,{N}} y(\bm{x}_i) \,.
\end{equation}

\red{The article \blue{by} \citet{PicGinRic2013} and the references therein discuss extensions of 
EGO to sequentially identify an $\bm{x} \in \arg\min y(\bf{x})$
when $y(\bm{x})$ \blue{observations contain measurement error,}
i.e., when the observed value at $\bm{x}$ is
$$
y^o(\bf{x}) = y(\bf{x}) + \epsilon(\bf{x}),
$$
where $\epsilon(\bf{x})$ is a white noise process
with variance ${\tau^2}$. } 
In this case, various approximations of the unobserved $y_{\min}^n$ have been suggested, including the standard plugin approach where $y_{\min}^n$ is approximated by $y_{\min}^n = \min_{i=1,\ldots,n} \hat{y}(\bm{x}_i)$, see \citet{PicGinRic2013} and references therein. Here we approximate $y_{\min}^n$ by $\min_{i=1,\ldots,n} \hat{y}(\bm{x}_i) - 2 \tau$, where $\tau$ is the noise standard deviation. Indeed, decreasing  $y_{\min}^n$ in \eqref{eqEXPIMP} typically increases the value of the expected improvement at input points with large uncertainties and large predictions, compared to points with small uncertainties and small predictions. Hence, this promotes exploration. Furthermore, the choice of the factor $-2$ is consistent with the common practice in nuclear safety of penalizing Monte Carlo simulation results by taking $5 \%$ or $95 \%$ Gaussian quantiles.

In this paper the 
problem of maximizing $y(\bm{x})$ over $\bm{x}$ in an unknown ${\cal X}$
is solved by identifying  ${\cal X}$ using one of the two methods described in
Section~\ref{sec:computation}.  Then the problem
\begin{equation} 
\label{eq:global:optim:expert:knowledge}
\bm{x}^\star
\in 
\argmax_{
\bm{x} \in {\cal X}
}
y( \bm{x} )
\end{equation}
is solved by the stochastic version of the EGO algorithm. Because 
 the empirically determined optimization domains for both examples in 
 Section~\ref{sec:examples} are subsets of hyperplanes, 
 the following adjustment is  made.  The  EGO algorithm is applied to maximize  the 
 expected improvement over $\bm{x} \in E$
where $E = \bm{\Lambda}{\cal X}$ and the  linear
transformation  $\bm{\Lambda}$  is stated 
in \citet{LoeWilMoo2013}.  The dimension of $E$ is 
one less than the number of components of $\bm{x} \in {\cal X}$.


\section{{Empirical Determination of the Input Domain}} 
\label{sec:computation}

This section describes two methods for identifying  
{a set} of positive input curves $\bm{x} = (x_1,\ldots, x_d)^\top$ 
which satisfy $\sum_{j=1}^{d} x_{j} =d$
and that are ``near'' the historical set of curves,
$\bm{x}_{i} = \left(x_{i,1},\ldots,x_{i,d} \right)^\top$, $i = 1,\ldots,n$.
\red{These constructions recognize that the historical curves form
a skeleton  of the total set of
curves that should be considered as the input domain for the 
optimization problems considered in this
paper.  
Informally, we use
the notation $\mathcal{X}$ to denote the input space.  
The first approach introduced in this section defines $\mathcal{X}$ by
constraints based on a mixture of expert knowledge of the physical system
and/or  graphical analysis   
of $\bm{x}_1,\ldots,\bm{x}_n$.
The second approach defines $\mathcal{X}$ as a  kernel density 
estimate formed from 
the coefficients of the projections of the
historical $\bm{x}_{i}$
onto a basis of spline functions.}

\subsection{Defining $\mathcal{X}$
Using Expert Knowledge and/or Empirical Experience}
\label{subsec:expert}

\blue{The first approach identifies}
$\mathcal{X}$ to be positive $\bm{x}$ curves 
using constraints determined by expert knowledge
and/or empirical experience.  The latter uses a visual 
 analysis of the $n$ historical curves.
As an example, the following four 
constraints based on the historical data were \blue{used}
 in the examples of Section~\ref{sec:examples}. 

\noindent
1.  {\bf Bound Constraints} at each \blue{of the $d$
component positions} of $\bm{x}$ 
\begin{equation}
\label{constraint1}
\min_{i=1,\ldots,n} ({x}_{i,j}) - \epsilon \leq  x_j \leq 
\max_{i=1,\ldots,n} ({x}_{i,j})  + \epsilon 
\end{equation}
where $j \in \{ 1,\ldots,d \}$ and
$\epsilon  \geq 0$ is a user-specified  
tolerance level.

\noindent
2.  {\bf Bounds on Incremental Changes} in consecutive 
components of $\bm{x}$
\begin{equation}
\label{constraint2}
\min_{i=1,\ldots,n} 
\left[ {x}_{i,j+1} - {x}_{i,j} \right] 
- \epsilon \leq  x_{j+1} - x_j 
\leq 
\max_{i=1,\ldots,n} \left[ 
{x}_{i,j+1} - {x}_{i,j} \right] + \epsilon 
\end{equation}
for all $j \in \{ 1,\ldots,d-1 \}$ where $\epsilon >0$
is a user-specified  tolerance level.

\noindent
3.  {\bf Constraints on Maximum Variation of $\bm{x}$}
\begin{equation}
\label{constraint3}
\max_{j=j_1,\ldots,j_2}| x_{j+1} - x_j | 
\leq
\max_{i=1,\ldots,n}
\max_{j=j_1,\ldots,j_2}|{x}_{i,j+1} - {x}_{i,j} |
+ \epsilon,
\end{equation}
where $1 \leq j_1 < j_2 \leq d$ and \blue{$\epsilon  \geq 0$} are user-specified.

\noindent
4.  {\bf Constraints on 
Maximum Total Variation of $\bm{x}$}
\begin{equation}
\label{constraint4}
\sum_{j=j_1}^{j_2} |x_{j+1} - x_j | 
\leq
\max_{i=1,\ldots,n}
\sum_{j=j_1}^{j_2} |{x}_{i,j+1} - {x}_{i,j} |
+ \epsilon,
\end{equation}
where $1 \leq j_1 < j_2 \leq d$ and \blue{$\epsilon  \geq 0$} are user-specified.
In other cases, more general linear or non-linear 
constraints such as
$$
\bm{A} 
\begin{bmatrix}
\bm{x}_i \\
  \bm{x}
\end{bmatrix}
 \blue{\leq} \  \bm{b}
$$
for  $i = 1,\ldots,n$, or
\[
\begin{bmatrix}
f_{1}(\bm{x}_1,\ldots,\bm{x}_n,\bm{x})
\\
\vdots \\
f_{s}(\bm{x}_1,\ldots,\bm{x}_n,\bm{x})
\end{bmatrix}
\leq 
\begin{bmatrix}
b_1 \\
\vdots \\
b_s
\end{bmatrix}
\]
could be used. We remark that the contraints 1 and 2 above are linear with respect to $\bm{x}$, while the constraints 3 and 4 are non-linear with respect to $\bm{x}$.


\subsection{Defining 
$\mathcal{X}$ by Projections onto a Basis Set}

\label{subsec:kernel}


\noindent
{\bf Projecting $x(t)$ onto the Set of 
Spline Basis Functions}
\label{subsubsection:splines}

\begin{figure} 
\begin{center}
\includegraphics[width=10cm]{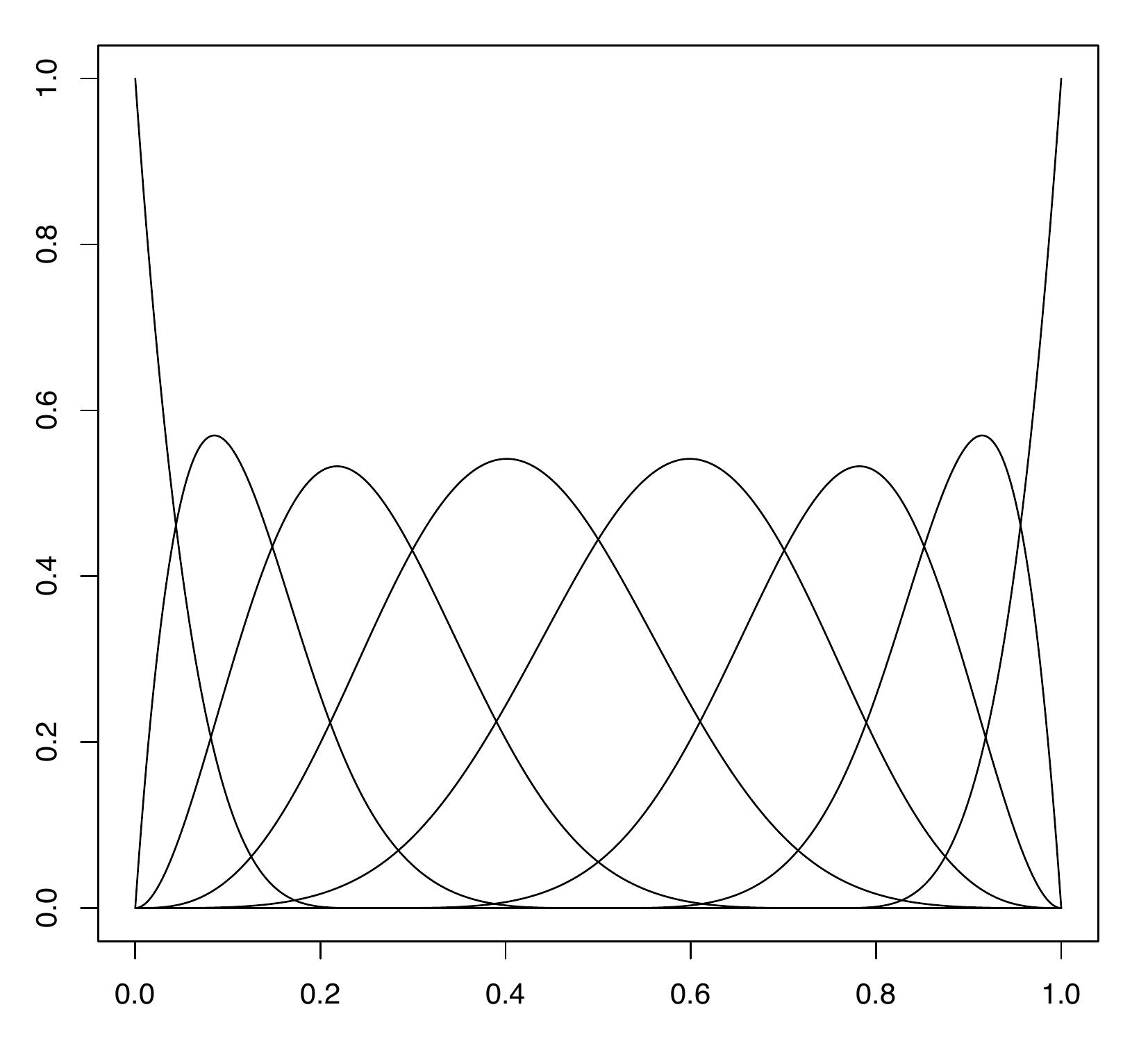}
\end{center}
\caption{$K =8$ spline functions of order $5$;
each spline can be identified by the location of its maximum value.}
\label{fig:spline:func}
\end{figure}

The references
\cite{ramsay2006functional} and \cite{muehlenstaedt2017computer}
provide an introduction to spline basis functions.
Let $\mathbb{N}$ denote the set of positive integers.
Briefly, a spline basis of order $m$, 
$m \in \mathbb{N}$, is a 
set of functions 
$B_{i,m} : [0,1] \to \mathbb{R}^+$, for 
$i =1, \ldots , K$ where $K \in \mathbb{N}$ is the number of 
spline functions.  Here $m$ is 
called the order of the spline.
{Figure \ref{fig:spline:func} illustrates the notation.}

The {\em projection} of a given positive real-valued
function $x$$:$ $[0,1] \to \mathbb{R}^+$ onto 
$\{B_{i,m}(t)\}_{i=1}^{K}$ is the function 
\[
\widehat{x}^{(m,K)}
= 
\widehat{x}^{(m,K)}(t) = 
\frac{ 
\widetilde{x}^{(m,K)}(t) 
}
{
 (1/d) \sum_{j=1}^d \widetilde{x}^{(m,K)} (t_j)
 }
\]
where 
\begin{equation} 
\label{eq:alpha:star}
 \widetilde{x}^{(m,K)}(t)  
 = \sum_{k=1}^K \alpha^\star_k B_{k,m}(t),\ 
 \ \mbox{and} \ 
(\alpha^\star_1,\ldots,\alpha^\star_K)
\in \argmin_{\alpha \in \mathbb{R}^K}
\int_{0}^1
\left(
x(t) -  \sum_{i=1}^K \alpha_i B_{i,m}(t)
\right)^2
dt
\end{equation}
which shows that $\widehat{x}^{(m,K)}$
has the form 
\begin{equation} 
\label{eq:alpha:spline}
\widehat{x}^{(m,K)}(t)
=
\sum_{k=1}^K \widehat{\alpha}_k B_{k,m}(t).
\end{equation}

The coefficients $\bm{\alpha}^\star = 
(\alpha^\star_1,\ldots,\alpha^\star_K)$ have an explicit expression as the least square solution to (\ref{eq:alpha:star}) and 
hence  ${\widehat{\bm{\alpha}}} = ( \widehat{\alpha}_1,\ldots, 
\widehat{\alpha}_K)$ is straightforward to obtain.

Recall that the observed data for the $i^{th}$ curve
is the vector 
$\bm{x}_i = (x_i(t_1), \ldots, x_i(t_d))^\top$, $i = 1, \ldots, n$.
To apply (\ref{eq:alpha:star}) to the function 
$\bm{x} = \bm{x}_i$,
let $\phi_{i}$ denote the 
spline interpolating function 
satisfying $\phi_i(t_1) = x_i(t_1),\dots,\phi_i(t_d) = x_i(t_d)$.
Then $\bm{\alpha}^\star$ corresponding
to $\bm{x}_i$ is obtained from \eqref{eq:alpha:star} by replacing
 $x(t)$  by $\phi_i(t)$.
(In the Section~\ref{sec:examples} examples, $\phi_i(t)$
is obtained by the R function 
\url{splinefun} in the package \url{splines}.)

\medskip

\noindent
{\bf Kernel Density Estimation}
\label{subsubsection:kde}

Let $\widehat{\bm{\alpha}}^{(i)} = 
\left( \widehat{\alpha}^{(i)}_1, \ldots,
\widehat{\alpha}^{(i)}_K \right)$ 
denote the $(\widehat{\alpha}_1,\ldots$,$\widehat{\alpha}_K)$ in \eqref{eq:alpha:spline} for the $i^{th}$ input curve
$\bm{x}_i$, $i=1,\ldots,n$.
Consider 
the following kernel density estimation procedure based on 
the set 
$\left\{ \widehat{\bm{\alpha}}^{(1)}, \ldots \right.$,  
$\left. \widehat{\bm{\alpha}}^{(n)} \right\}$.
Following the approach of \cite{perrin2018data}, let 
$\phi(\bm{\cdot})$ denote the 
probability density function of the univariate
standard Normal distribution. Given $K$ and 
positive scale factors $\bm{\lambda}= 
\left( \lambda_1,\ldots,\lambda_K \right)$, let 
\begin{equation} 
\label{eq:f:lambda}
\rho_{\lambda_1,\ldots,\lambda_K} (\bm{\alpha})
=
\rho_{\bm{\lambda}} (\bm{\alpha})
=
\frac{1}{n}
\sum_{i=1}^n
\prod_{k=1}^K
\frac{1}{\lambda_k}
\phi \left(
\frac{ \alpha_k - \widehat{\alpha}^{(i)}_k }{ \lambda_k }
\right)
\end{equation}
define a function from $\mathbb{R}^K$ to $\mathbb{R}^+$
where $\bm{\alpha} = (\alpha_1, \ldots,\alpha_K)$.
It is straightforward to check that $\rho_{\bm{\lambda}} 
(\bm{\alpha})$
has integral one over $\mathbb{R}^K$. 
Intuition suggests that given a scaling 
$\bm{\lambda} \in (0,\infty)^K$,
$\rho_{\bm{\lambda}} (\bm{\alpha})$ 
is {\em large} for choices of $\bm{\alpha}$ that
are compatible with the {\em set of observed input curves}. 

In this paper the 
scale parameters $\lambda_1,\ldots,\lambda_K$  
are selected 
by cross validation using
\begin{equation}
\label{eq:compat_lambda}
\widehat{\bm{\lambda}} 
= 
(\widehat{\lambda}_1,\ldots,\widehat{\lambda}_K)^\top
\in \argmax_{(\lambda_1,\ldots,\lambda_K) \in (0,\infty)^K}
\sum_{i=1}^n
\log 
\left(
\rho^{-i}_{\bm{\lambda}} (\widehat{\bm{\alpha}}^{(i)})
\right)
\end{equation}
where $\rho^{-i}_{\bm{\lambda}}$ 
is obtained from \eqref{eq:f:lambda} by removing 
$\widehat{\bm{\alpha}}^{(i)}$ from the
set 
$\left\{ \widehat{\bm{\alpha}}^{(i)} \right\}_{i=1}^{n}$
(and decrementing $n$ to $n-1$).
Thus $\rho_{\widehat{\bm{\lambda}}}(\bm{\alpha})$ 
can be viewed as a kernel density estimator
most compatible with the coefficients 
$\left\{ \widehat{\bm{\alpha}}^{(1)}, \ldots \right.$,  
$\left. \widehat{\bm{\alpha}}^{(n)} \right\}$ from 
$\bm{x}_1$, \ldots, $\bm{x}_n$.
Thus the value of 
$\rho_{\widehat{\bm{\lambda}}}(\bm{\alpha})$ 
is used to 
quantify the level of ``realism'' of curves having form 
$\sum_{k=1}^K \alpha_k B_{k,m}$ to the  observed 
$\bm{x}_1$, \ldots $\bm{x}_n$.

\medskip

\noindent
{\bf Threshold Selection}
\label{subsubsection:threshold}

In the following, 
$\widehat{\rho}(\bm{\alpha}) =
\rho_{\widehat{\bm{\lambda}}}(\bm{\alpha})$ 
denotes the estimated compatibility function
in (\ref{eq:f:lambda}) and (\ref{eq:compat_lambda}).
To select discretized curves  $\bm{x}$ most 
compatible with the training data, 
we choose $T>0$ such that $\bm{\alpha} \in \mathbb{R}^K$ is considered compatible with
$\bm{x}_1, \ldots, \bm{x}_n$  
if and only if $\widehat{\rho}( \bm{\alpha} ) \geq T$.

The value $T$ is chosen as follows.   
For $\bm{\alpha} \in \mathbb{R}^K$,
let $x_{\bm{\alpha}}(t) = 
 \sum_{k=1}^K {\alpha}_k B_{k,m}(t)
$; given $\Delta~>~0$, select a $T>0$ such  that
any $\bm{\alpha}$ which satisfies
\begin{equation}
\label{eq:4.delta}
\left( \int_0^1 \left[ x_{\bm{\alpha}}(t) 
- 
x_{\widehat{\bm{\alpha}}^{(i)}}(t) 
\right]^2 dt \right)^{1/2} 
\leq \Delta 
\end{equation}
for at least one $i \in \{1,\ldots,n\}$
 also satisfies 
$\widehat{\rho}(\bm{\alpha}) \geq T$.  Inspection of \eqref{eq:f:lambda} 
shows that $\widehat{\rho}(\bm{\alpha}) \geq T$ 
holds provided, for some $i \in \{1 , \ldots , n \}$,
\begin{equation} 
\label{eq:pdf:one:isolated:point}
\frac{1}{n}
\prod_{k=1}^K
\frac{1}{\widehat{\lambda}_k}
\phi \left(
\frac{ \alpha_k - 
\widehat{\alpha}^{(i)}_k }{ \widehat{\lambda}_k }
\right) \geq T
\end{equation}
and (\ref{eq:4.delta}) holds 
for this $\bm{\alpha}$ and $i$.
The expression on the right hand side of 
\eqref{eq:pdf:one:isolated:point} is the limiting value of \eqref{eq:f:lambda} corresponding to 
the case where $\widehat{\bm{\alpha}}^{(i)}$ 
is infinitely distant from all
$\{ \widehat{\bm{\alpha}}^{(\ell)}\}_{\ell \neq i}$ 
and where 
$\left( \int_0^1 \left[ 
x_{\bm{\alpha}}(t) 
- 
x_{\widehat{\bm{\alpha}}^{i}}(t)
\right]^2 dt \right)^{1/2} \leq \Delta $.
Based on the above observations, the selected threshold 
$\widehat{T}$ is defined as
\begin{equation}
\label{eq:T}
\widehat{T}
=
\min_{ \substack{
\bm{\alpha} \in \mathbb{R}^K
\\
\left( \int_0^1 x_{\bm{\alpha}}(t)^2 dt \right)^{1/2} \leq \Delta 
} }
\frac{1}{n}
\prod_{k=1}^K
\frac{1}{\widehat{\lambda}_k}
\phi \left(
\frac{ \alpha_k }{ \widehat{\lambda}_k }
\right).
\end{equation}
In practice, the calculation of $\widehat{T}$ in
(\ref{eq:T}) is 
straightforward since one can precompute the
Gram matrix with
$(i,j)^{th}$ element
$\int_0^1 B_{i,m} (t) B_{j,m} (t) dt$.

Let $\bm{x}(\bm{\alpha})$ denote the $d \times 1$ vector with
$j^{th}$ element
$\left[ \sum_{k=1}^K \alpha_k B_{k,m}(t_j) \right]$
for $j=1,\ldots,d$. 
As for the historical data, scaling $\bm{x}(\bm{\alpha})$ 
by the average of its components, i.e., by
$\bar{x}(\bm{\alpha}) = 
(x_1(\bm{\alpha})+\ldots+x_d(\bm{\alpha}))/d$
results in a positive point on the $d$-hyperplane
(when $\bm{\alpha}$ has positive components).
To select $\bm{\alpha}$ compatible
with $\widehat{\rho}(\bm{\alpha}) > T$, 
compute
\begin{equation} 
\label{eq:global:optimization:kde}
\bm{\alpha}^*
\in 
\argmax_{  
\substack{
\bm{\alpha} \in [0,\infty)^K
\\
\widehat{\rho}(  \bm{x}(\bm{\alpha})  / \bar{x}(\bm{\alpha})   )
\geq \Delta
\\
}
}
y( \bm{x}(\bm{\alpha}) /\bar{x}(\bm{\alpha})  ).
\end{equation}
The 
optimization problem corresponds to minimizing a  
function where the constraints can be tested with negligible cost. 
This optimization takes place 
in the $K$-dimensional
space of the $\bm{\alpha}'s$.


\section{Worked Examples} 
\label{sec:examples}

The rationale behind our approaches to forming an 
optimization domain, i.e., using expert-determined constraints or projections, 
relies on the availability of a sufficient amount of relevant data.
Following an explanatory data analysis (EDA) of the burn-up profile data in Section~\ref{sec:EDA}, 
 Section~\ref{section:determining:admissible} applies both methods 
to identify a set of burn-up rate curves $\bm{x}$
that are consistent with those of spent 
fuel rods from nuclear plants based of the ``historical'' curves in the
axial burn-up profile database for pressurized water 
reactors available through OECD Nuclear Energy Agency Data Bank 
\citep{pwrbup1997}.

Subsection~\ref{section:optim:analytical} applies
the optimization method of Section \ref{sec:sequentialoptimization}
to a simple
analytic function where the answer and the 
performance of the 
optimization procedure is straightforward to understand.  
\blue{Then Subsection~\ref{subsec:optim:burnup} considers the
example introduced in Section \ref{sec:MathDesc} to
maximize the criticality coefficient for spent fuel rods. 
This process is termed Burn-up Credit Penalization in the nuclear industry.}
Finally, Subsection \ref{subsection:general example} discusses the generalizability of the methodology of this paper.
It  presents an additional analytical example 
where both the historical curves and the objective function are unknown.

\subsection{Exploratory Analysis of the Fuel Rod Data}
\label{sec:EDA}

\begin{figure}
\begin{center}
\includegraphics[width=8cm]{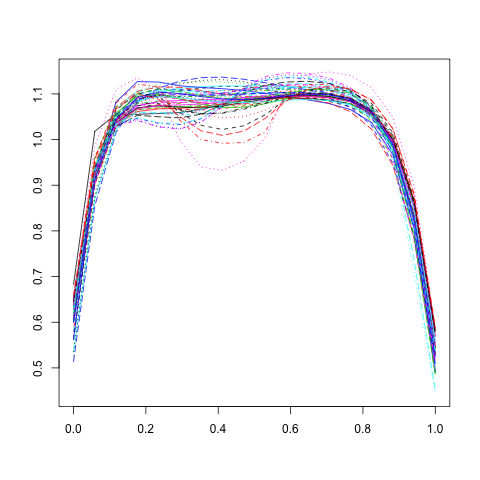}
\end{center}

\vspace{-.5in}

\caption{
Fifty representative burn-up rate curves from 
the population of 3,158 discretized historical curves.
 }
\label{fig:courbes_101-150}
\end{figure}

In this application, there are $n =$ 3,158
discretized burn-up rate
curves $\bm{x}_1$, \ldots, $\bm{x}_{3158}$,
each of which has been measured at the (same)
$d=18$ vertical measurement points
$(0,1 / 17$,\ldots, $16/17,1)$. 
\red{Recall that the
curves have been normalized so that
$\sum_{j=1}^{18} x_{i,j} = 18$, for $i=1,\ldots,$ 3,158, or equivalently to 
have an average burn-up rate  equal to one.}
Figure~\ref{fig:courbes_101-150} shows $50$ representative
curves from  the population of 
curves; all $50$ curves show a 
common `vertical-horizontal-vertical' shape
which is true of the majority of curves. 
A small minority of the population have a
more complex character.

\begin{figure}
\begin{center}
\begin{tabular}{cc}
\includegraphics[width=8cm]{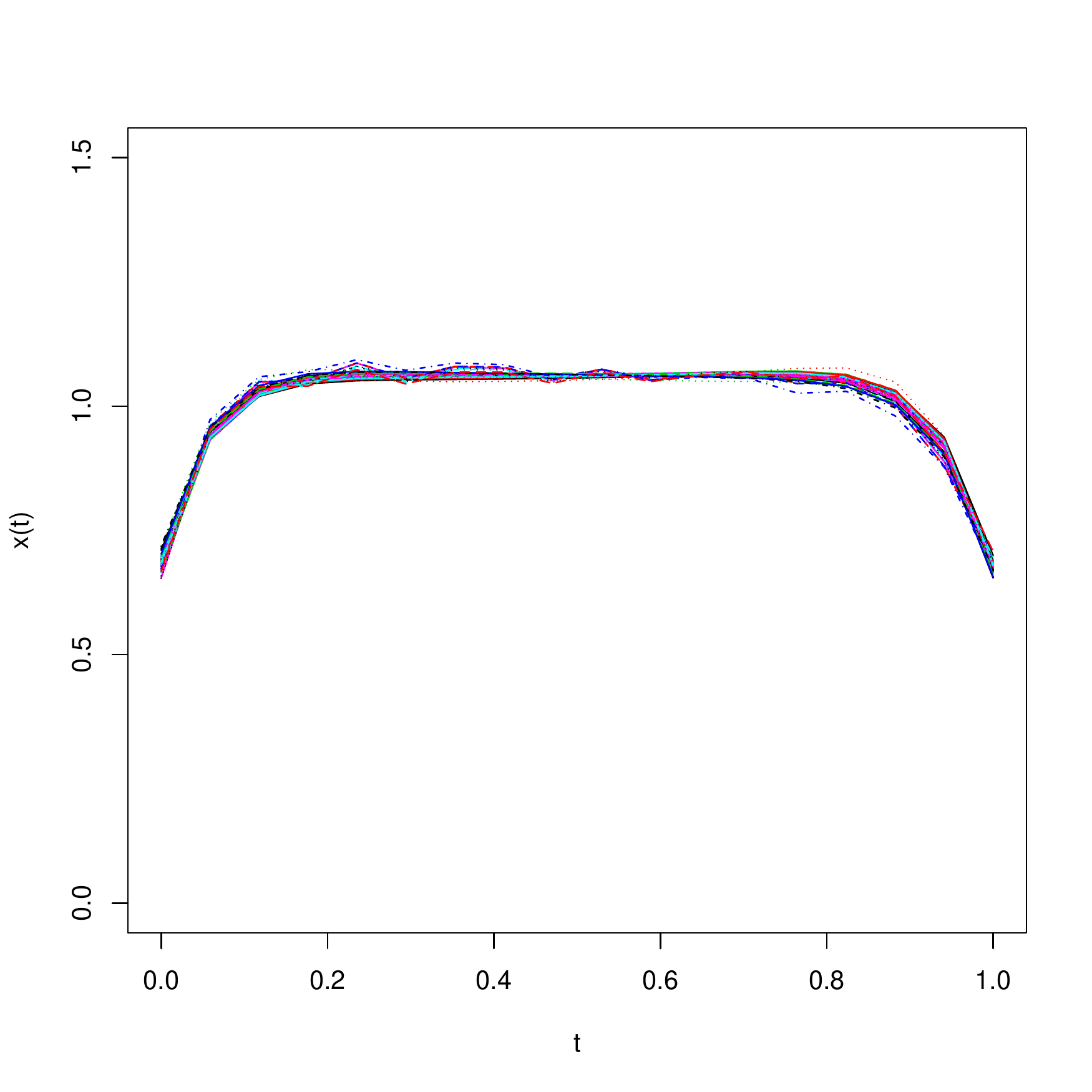}
&
\includegraphics[width=8cm]{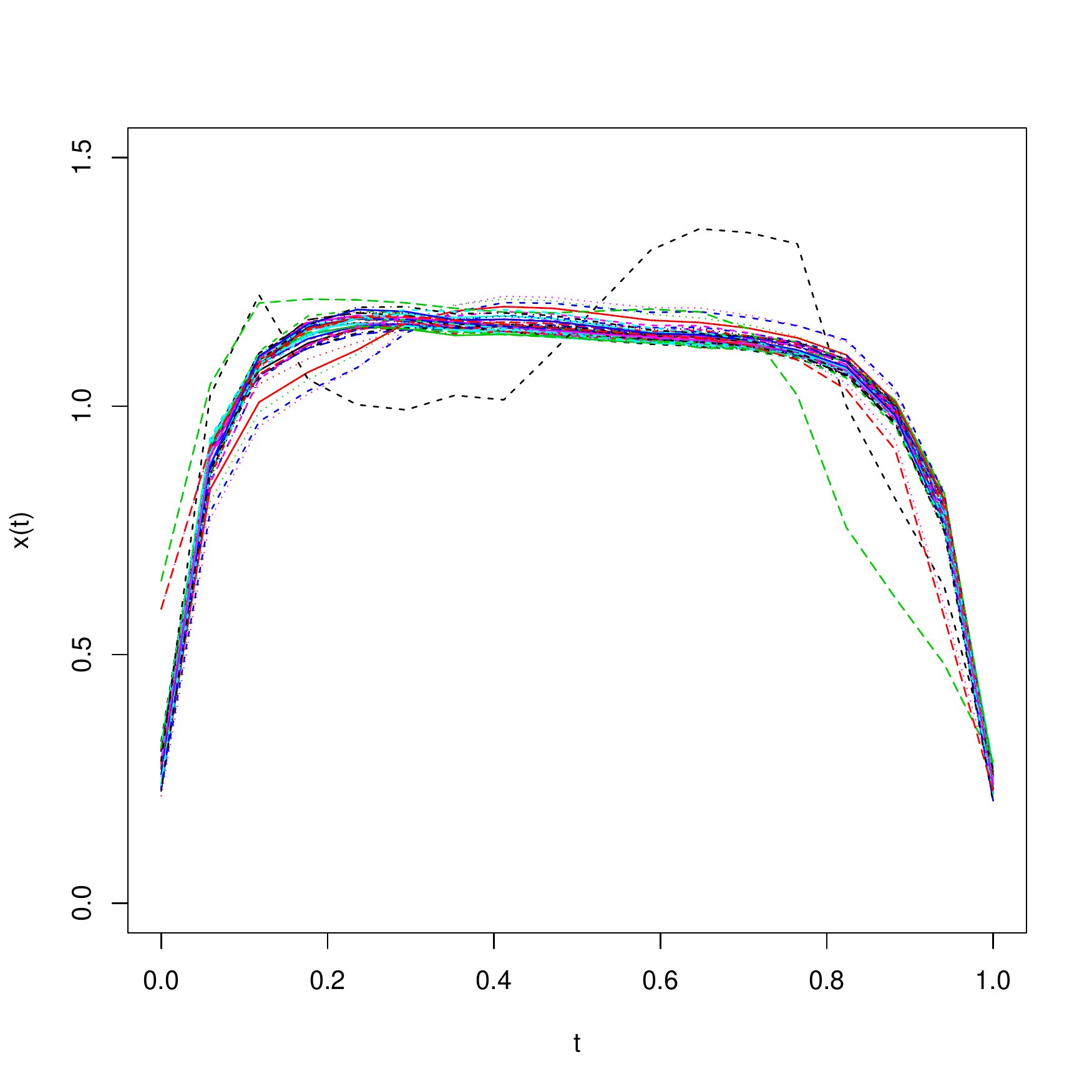}
\end{tabular}
\end{center}

\vspace{-.5in}

\caption{In the fuel rod application of Subsection~\protect\ref{sec:EDA},
the $50$ historical curves $\bm{x}_i$ with the smallest 
(left panel) and the largest (right panel) 
outputs $y(\bm{x}_i)$.}
\label{fig:X:lowest:highest:Y}
\end{figure}

\red{Because each run of the {\sl CRISTAL} code for this application
required only fifteen minutes, sufficient budget was 
available that  
the code was run for all 3,158 input functions.} 
Figure \ref{fig:X:lowest:highest:Y} plots the $50$ curves yielding 
the {\em lowest} values of the 
criticality coefficient, $y(\bm{x}_i)$ \blue{(between $0.86149$ and $0.86665$)},
and the $50$ curves yielding the {\em largest} values 
of the criticality coefficient \blue{(between $0.92758$ and $0.94123$)}. 
Visually, it is plain that rods which 
are evenly burned over $t$, i.e.,  which have constant
$x(t)$, are safest in the sense of having
small $y(\bm{x})$ values 
while rods that are burned unevenly  
are more hazardous. 

\subsection{Forming ${\cal X}$} 
\label{section:determining:admissible}
Section~\ref{sec:computation} described 
two methods for defining the domain 
of curves $\bm{x}$ having  
representative burn-up rates.  The first 
method combines graphical and numerical 
EDA with 
expert knowledge about the features of burned fuel rods; 
the second method selects $\bm{x}$ which are
``near'' to the body of basis representatives
of the original 3,158 curves. 
 

\medskip

\noindent
{\bf Defining ${\cal X}$ by EDA and  Expert-Type Constraints}

Based 
on visualization of the curves in
Figure~\ref{fig:courbes_101-150}, curves that satisfy
the following constraints are considered to 
have representative burn-up rates.

\begin{itemize}
\itemsep=.001in
\item {\bf Bound Constraints} in (\ref{constraint1}):
set $\epsilon = 0.05$
for the values of the first 
and last measured burn-up rate, i.e.,
$x_{i,1}$ and $x_{i,18}$;

\item  
{\bf Bounds on Incremental Changes}  in (\ref{constraint2}):
set $\epsilon = 0.03$ for each
of the increments $|x_{i,j+1} - x_{i,j}|$, 
$j=1, 2, 16$, and $17$;

\item 
{\bf Constraints on Maximum Variation of $\bm{x}$} in (\ref{constraint3}):
set $\epsilon = 0.03$, $j_1=3$, and $j_2=16$;

\item 
{\bf Constraints on 
Maximum Total Variation of $\bm{x}$} in (\ref{constraint4}):
set $\epsilon = 0.1$,
$j_1=3$, and $j_2=16$.
\end{itemize} 
All 3,158 historical curves satisfy these four constraints
{by definition}
and are thus part of the ${\cal X}$ domain defined by this criterion.

We remark that the choice of the above constraints only required basic physical knowledge, and mostly relied on a visual analysis of the historical curves. More precisely, it was observed that the curves have a ``vertical-horizontal-vertical'' pattern which yielded the two first constraints. It was also observed that the curves have moderate increments, which yielded the two last constraints.
This is a benefit of this methodology for determining the domain $\mathcal{X}$, since it is hence available to statisticians, even though the constraints could benefit from nuclear scientists' confirmation, particularly if the historical data base is not as rich as is the case here ($3, 158$ curves). In any cases, the methodology offers the possibility to benefit from expert knowledge, where an expert can suggest constraints with no knowledge of the historical curves. This opportunity is not taken here, but is a further potential benefit of the methodology.

\medskip

\noindent
{\bf Defining ${\cal X}$
as a Kernel Density Estimate}

\begin{figure}
\begin{center}
\includegraphics[width=8cm]{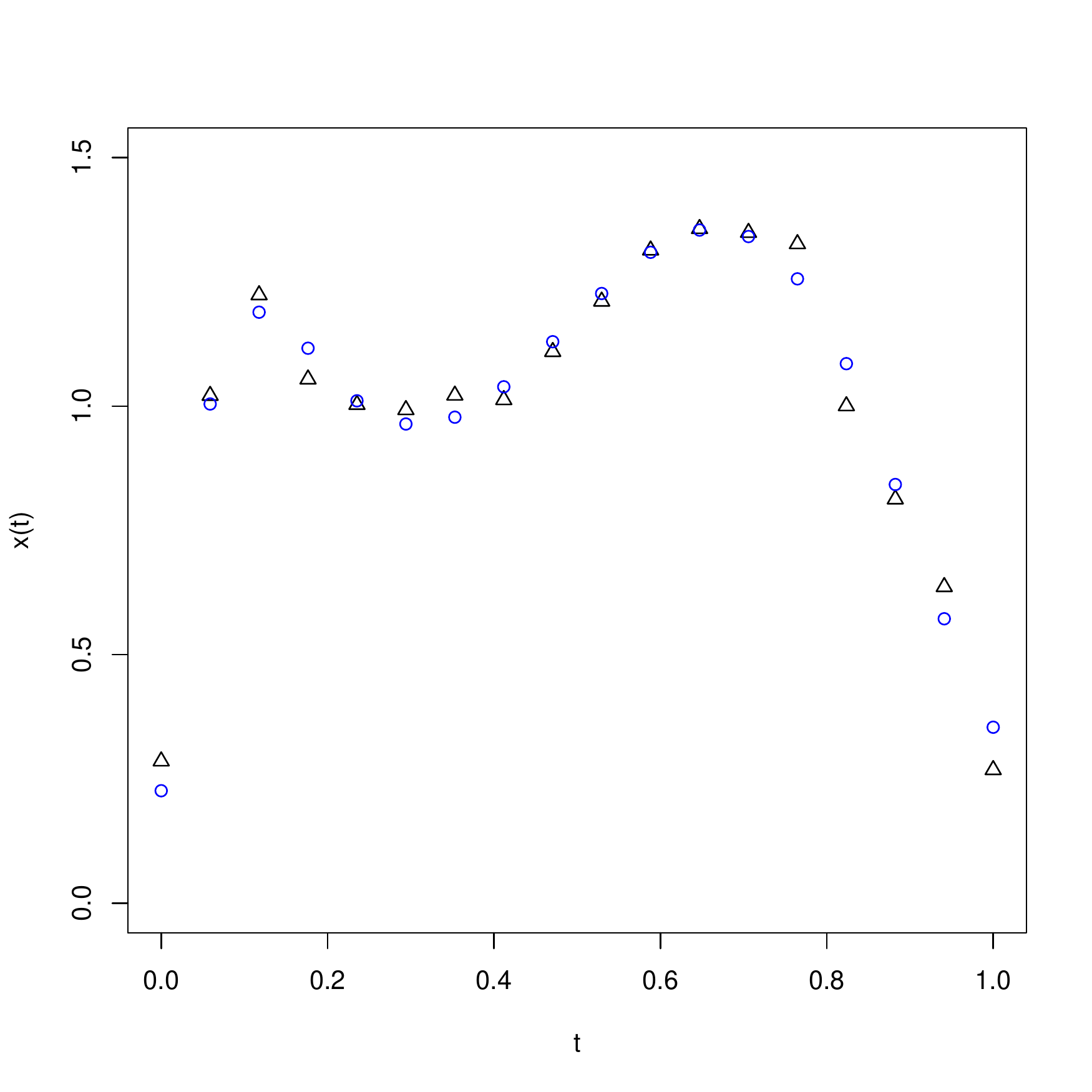}
\end{center}

\vspace{-.4in}

\caption{Original (black triangles) and spline approximation (blue circles) 
for the curve 
$\bm{x}_i = (x_{i,1}, \ldots,  x_{i,18})$ \blue{having the}
largest mean square difference 
compared with its spline approximation, among the 3,158 
historical curves.}
\label{fig:spline:worst:MSE}
\end{figure}

This application of Subsection~\ref{subsec:kernel}
takes $K = 8$ spline functions
of  order
$m=5$ (and are constructed 
using the R package \url{splines} 
with knot sequence $(0,0,0,0,0,0.25,0.5,0.75,1,1,1,1,1)$ and option \url{monoH.FC}). Figure \ref{fig:spline:func}
plots the resulting
set of spline
functions $\{B_{k,5}(t) \}_{k=1}^{8}$ over $t \in [0,1]$.
As described in Subsection \ref{subsubsection:splines}, each of the 3,158 
discretized curves $\bm{x}$ can be 
approximated by a spline  
$\sum_{k=1}^8 \widehat{\alpha}^{(i)}_k B_{k,5}(t)$.  
This representation results in a dimension 
reduction from $18$ to $8$, and provides a good fit of the 
3,158 curves. Figure~\ref{fig:spline:worst:MSE} plots 
the original and spline approximation for the curve 
$\bm{x}_i$ 
having {\em largest mean square difference} from
its spline approximation, among the 3,158 historical curves. 

\begin{figure}
\begin{center}
\includegraphics[width=8cm]{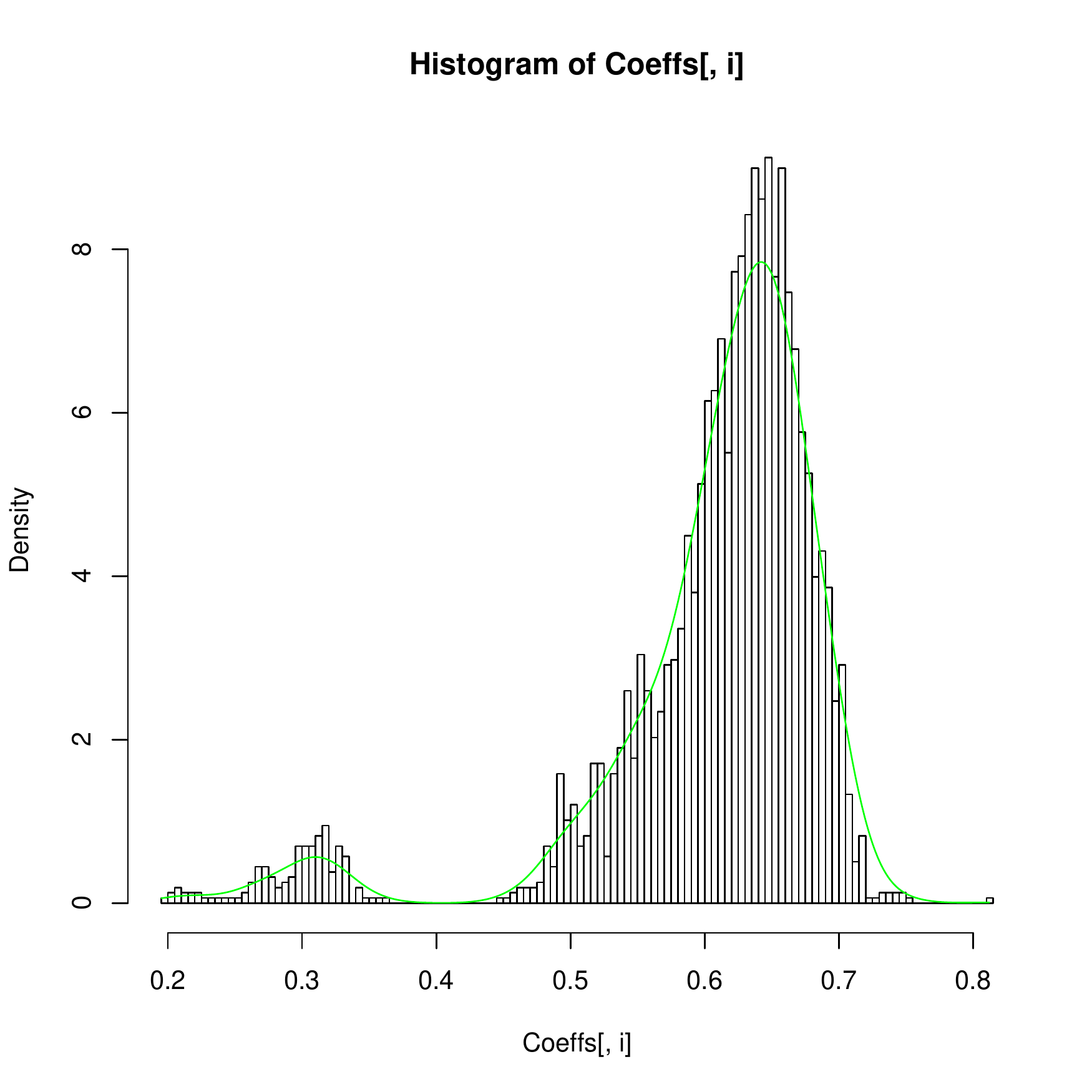}
\end{center}

\vspace{-.4in}

\caption{Plot of the first marginal probability density function of the 
($8$-d) probability density function 
$\rho_{\widehat{\lambda}_1,\ldots,\widehat{\lambda}_8}$ (in green), together with the histogram of the first components 
{$\widehat{\alpha}^{(1)}_1,\ldots,\widehat{\alpha}^{(3,158)}_1$}
from the spline kernel basis 
representations of the 3,158 curves.}
\label{fig:hist:pdf}
\end{figure}

Kernel density estimation is performed 
as described in Subsection \ref{subsubsection:kde}. The window vector 
obtained is
$(\widehat{\lambda}_1,\ldots,\widehat{\lambda}_8 ) \approx  (0.018, 0.019, 0.019, 0.018, 0.017, 0.015, 0.021, 0.017)$. 
To illustrate, Figure \ref{fig:hist:pdf} plots the density
of the first marginal probability density function of the 
($8$-dimensional) probability density function 
$\rho_{\widehat{\lambda}_1,\ldots,\widehat{\lambda}_8}$ 
together with the histogram of 
{$\widehat{\alpha}^{(1)}_1,\ldots,\widehat{\alpha}^{(3,158)}_1$}.

\begin{figure}
\begin{center}
\includegraphics[width=8cm]{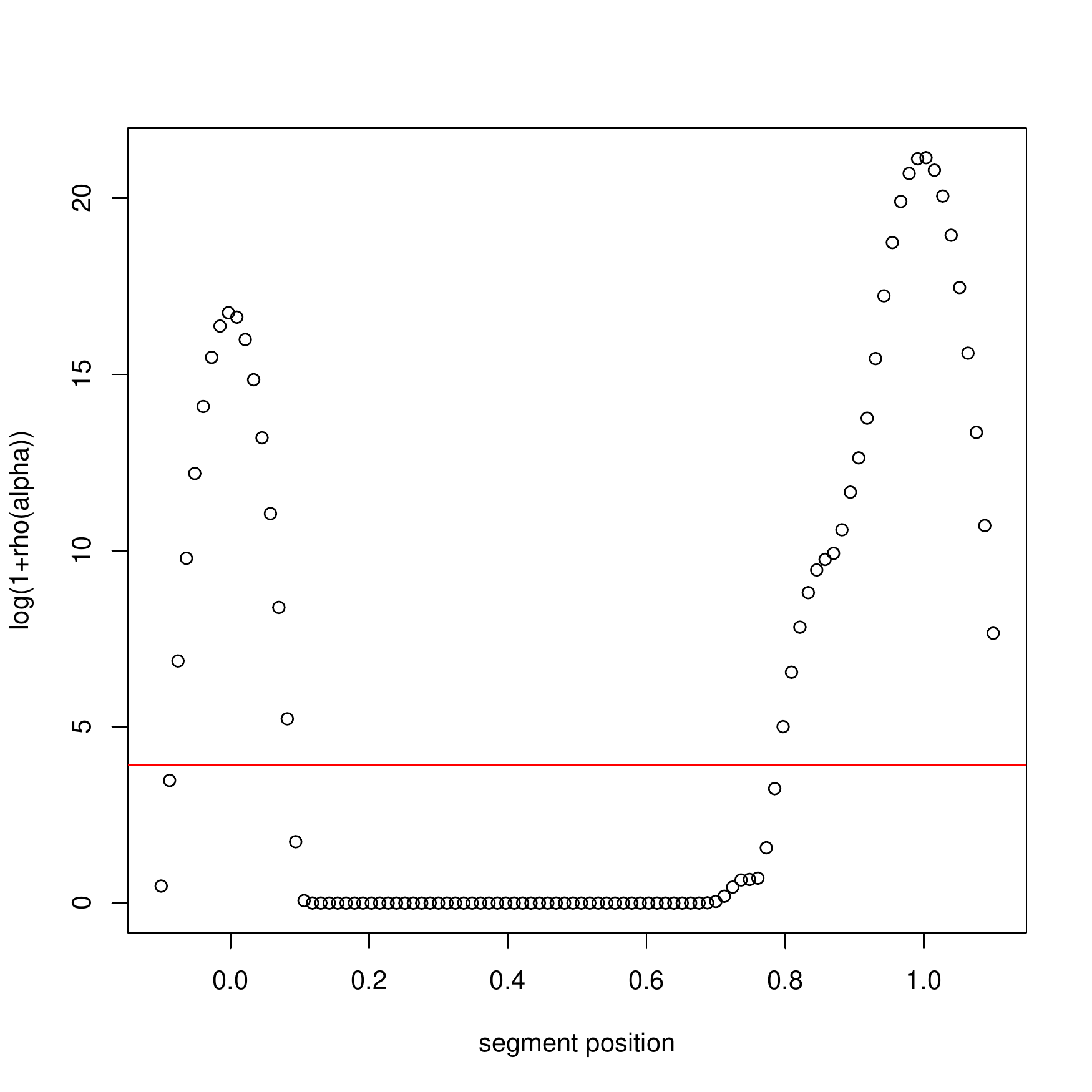}
\end{center}

\vspace{-.5in}

\caption{Values of $\log(1+\widehat{\rho}({\bm{\alpha}} ))$ (black dots), for 
$\bm{\alpha}$ belonging to the segment with endpoints {obtained from} 
a isolated historical coefficient vector and a non-isolated one. 
\blue{The red line is the logarithm of (1 plus the threshold). } }
\label{fig:illustration:threshold}
\end{figure}

The threshold value is selected as described in Subsection \ref{subsubsection:threshold}, where
{$\Delta = 0.05$} is chosen, and is
{$\widehat{T} =  54.86$}.
\red{Figure \ref{fig:illustration:threshold} provides a visual insight of the domain 
$\{ \bm{\alpha}$ $:$ $\widehat{\rho}(\bm{\alpha}) \geq \widehat{T} \}$.} 
\red{Coefficient vectors 
$\widehat{\bm{\alpha}}^{(i_1)}$ and $\widehat{\bm{\alpha}}^{(i_2)}$ are considered for two
of the historical inputs, where $\widehat{\bm{\alpha}}^{(i_1)}$ is numerically distant 
from the remaining $\{ \widehat{\bm{\alpha}}^{(i)} \}_{i \neq i_1}$, 
while $\widehat{\bm{\alpha}}^{(i_2)}$ has closer neighbors. As a consequence $\widehat{\rho}( \widehat{\bm{\alpha}}^{(i_2)} ) > \widehat{\rho}( \widehat{\bm{\alpha}}^{(i_1)} )$.} The value of 
$\widehat{\rho}(\bm{\alpha} )$ is plotted, for $\bm{\alpha}$ belonging to the 
segment with endpoints
{
$ \widehat{\bm{\alpha}}^{(i_1)}  - 0.1 \left( \widehat{\bm{\alpha}}^{(i_2)} - \widehat{\bm{\alpha}}^{(i_1)} \right) $ 
and 
$ \widehat{\bm{\alpha}}^{(i_1)}  + 1.1 \left( \widehat{\bm{\alpha}}^{(i_2)} - \widehat{\bm{\alpha}}^{(i_1)} \right) $}.
 One observes that the parts of the 
segment close to ${\widehat{\bm{\alpha}}^{(i_1)} }$ and 
${\widehat{\bm{\alpha}}^{(i_2)} }$ 
correspond to admissible $\bm{\alpha}$'s, while the middle of the segment 
corresponds to inadmissible $\bm{\alpha}$'s. Recall that $\Delta$ is user 
selected and that decreasing it \red{increases} the threshold and vice 
versa.

\subsection{Optimization of an Analytical Function} 
\label{section:optim:analytical}

This subsection maximizes an analytical function
$y(\bm{x})$ over the domain of the simplex 
determined from the $n =$  $3$,$158$ historical curves
from the nuclear power industry.  
Both the 
expert-type constraints  methodology
and the kernel density approximation will be illustrated to 
identify this input space.
The analytical function to be maximized is
\begin{equation} 
\label{eq:analytical:function}
y_{a}(\bm{x})
=
-
|| \bm{x} - \bm{x}_0 ||_2
-
\sin \left( 3 || \bm{x} - \bm{x}_0 ||_2  \right)^2,
\end{equation}
where $\bm{x}_0$ is a fixed
one of the $n =$ 3,158 historical curves and
$|| \bm{w} ||_2 = ( \bm{w}^\top  \bm{w} )^{1/2}$ for a column vector $\bm{w}$.
The unique maximizer of $y_{a}(\bm{x})$ in 
\eqref{eq:global:optim:expert:knowledge} is  $\bm{x} = \bm{x}_0$ with
optimal value $y_{a}(\bm{x}_0) = 0$. 
The function $y_a(\bm{x})$ is observed with additive 
Gaussian noise having mean zero and variance $0.0005^2$. 
This mild Monte Carlo noise in the $y(\bm{x})$ function
mimics that present in the second example. 

The goal is to assess whether the expected improvement algorithm 
is able to converge to the global maximizer  
 for both the expert-type constraint domain in 
 \eqref{eq:global:optim:expert:knowledge} or  the 
 domain defined using kernel density
 approximation in \eqref{eq:global:optimization:kde}. 



\medskip

\noindent
{\bf EI  Optimization of \eqref{eq:analytical:function}
over ${\cal X}$ Determined by
Expert-Type Constraints} 
\label{section:expert:analytical}

The set of training curves used for 
the analytic function consisted of
$100$ 
 curves selected by a space-filling design
among the 3,158 historical \red{input vectors}.
The maximum 
value of $y_a(\bm{x})$ among the training data 
is approximately $-0.75$. 
Then $50$ additional discretized curves were 
selected from ${\cal X}$ using the EI/expert knowledge procedure.
The maximum $y_a(\bm{x})$  
increased to approximately $-0.17$ using the $50$ additional curves. 

\begin{figure}[h!]
\begin{center}
\includegraphics[width=10cm]{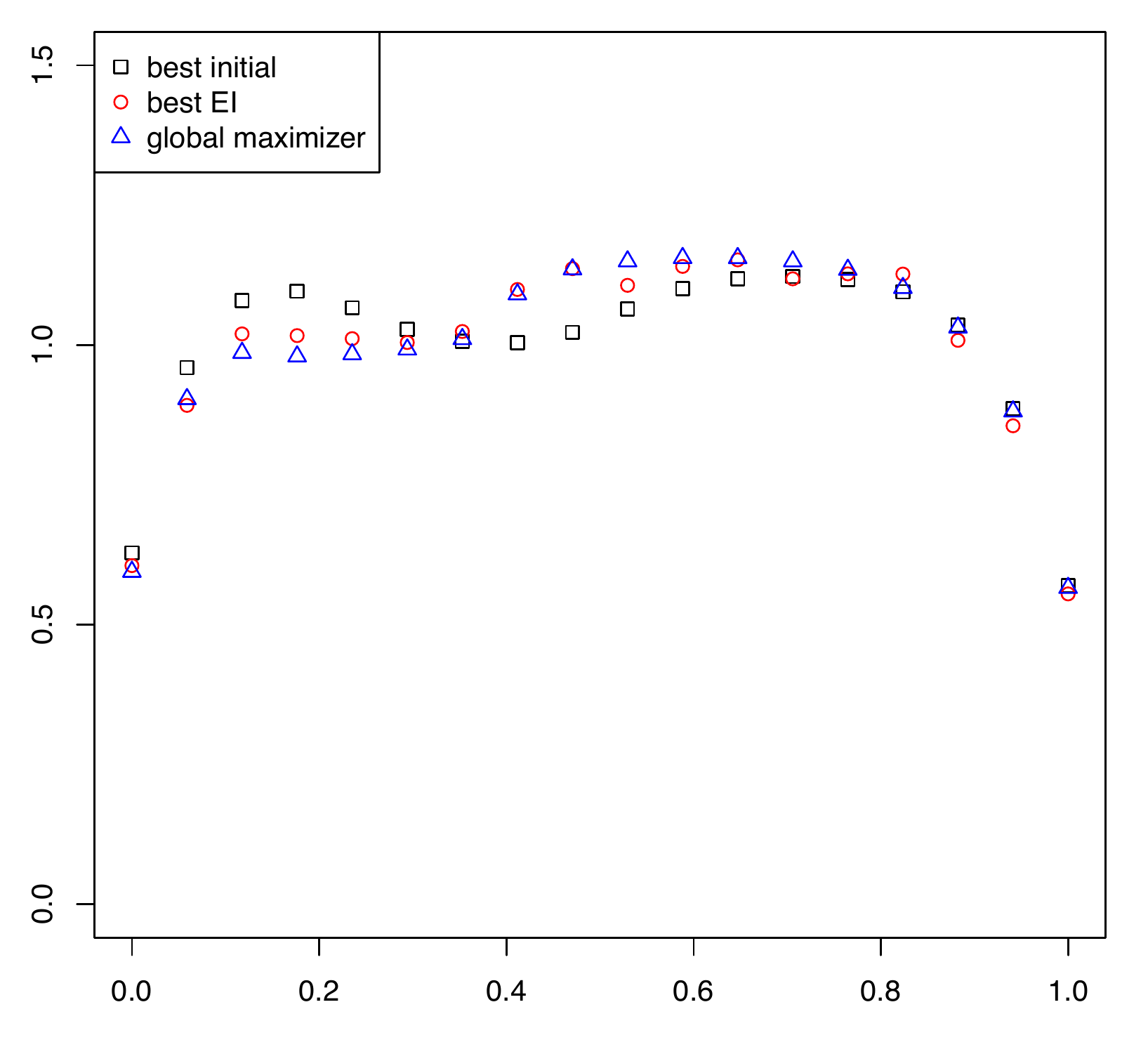}
\end{center}
\vspace{-.2in}
\caption{Three input curves for the 
analytic optimization problem
\eqref{eq:analytical:function}:
the true global $y_a(\bm{x})$  maximizer;
the $y_a(\bm{x})$ maximizer 
among the $100$ initial curves;  
the $y_a(\bm{x})$ maximizer 
among the $50$ curves added by EI/expert-type constraints based
on (\ref{constraint1})-(\ref{constraint4}).}
\label{fig:expert:knowledge:toy:function:curves}
\end{figure}

Figure~\ref{fig:expert:knowledge:toy:function:curves} 
provides a visual understanding 
of this performance by plotting 
three curves that illustrate the performance 
of the proposed procedure.  The first curve is $\bm{x}_0$ 
which denotes the true global maximizer; the second
curve, denoted
$\bm{x}_{\mathrm{init}}$, is the  maximizer of 
$y_a(\bm{x})$ 
among the $100$ initial curves; the third curve, denoted 
$\bm{x}_{\mathrm{EI}}$, is the  maximizer 
of $y_a(\bm{x})$ among the $50$ curves obtained by expected 
improvement. Observe that 
$\bm{x}_{\mathrm{EI}}$ is, visually, significantly closer to $\bm{x}_0$ 
than is $\bm{x}_{\mathrm{init}}$, 
which suggests the convergence of the 
expected improvement procedure. This example
also shows that the admissible set 
in \eqref{eq:global:optim:expert:knowledge} is amenable to maximization 
in practice. 

\noindent
{\bf EI Optimization of \eqref{eq:analytical:function}
over ${\cal X}$ Determined by
Kernel Density Estimation}

In this case, while the analytical function $y_a(\bm{x})$ is  
\eqref{eq:analytical:function}, 
the curve $\bm{x}_0$ is now 
given by $(x_0(t_1),\ldots,x_0(t_d))$, with 
\[
x_0(t)
=
\frac{
\sum_{i=1}^K \alpha_{0,i} B_{i,m}(t)
}{
\frac{1}{d}
\sum_{j=1}^d \sum_{i=1}^K \alpha_{0,i} B_{i,m}(t_j)
}
\]
where $\bm{\alpha}_0$ is one of the $3$,$158$ 
$\widehat{\bm{\alpha}}^{(1)},\ldots,\widehat{\bm{\alpha}}^{(3158)}$.
Thus in the optimization problem \eqref{eq:global:optimization:kde}, the global maximizer curve is given by $\bm{\alpha}^\star = \bm{\alpha}_0$. Noisy observations of $y_a(\bm{x})$ are obtained as above. 

\begin{figure}[h!]
\begin{center}
\includegraphics[width=10cm]{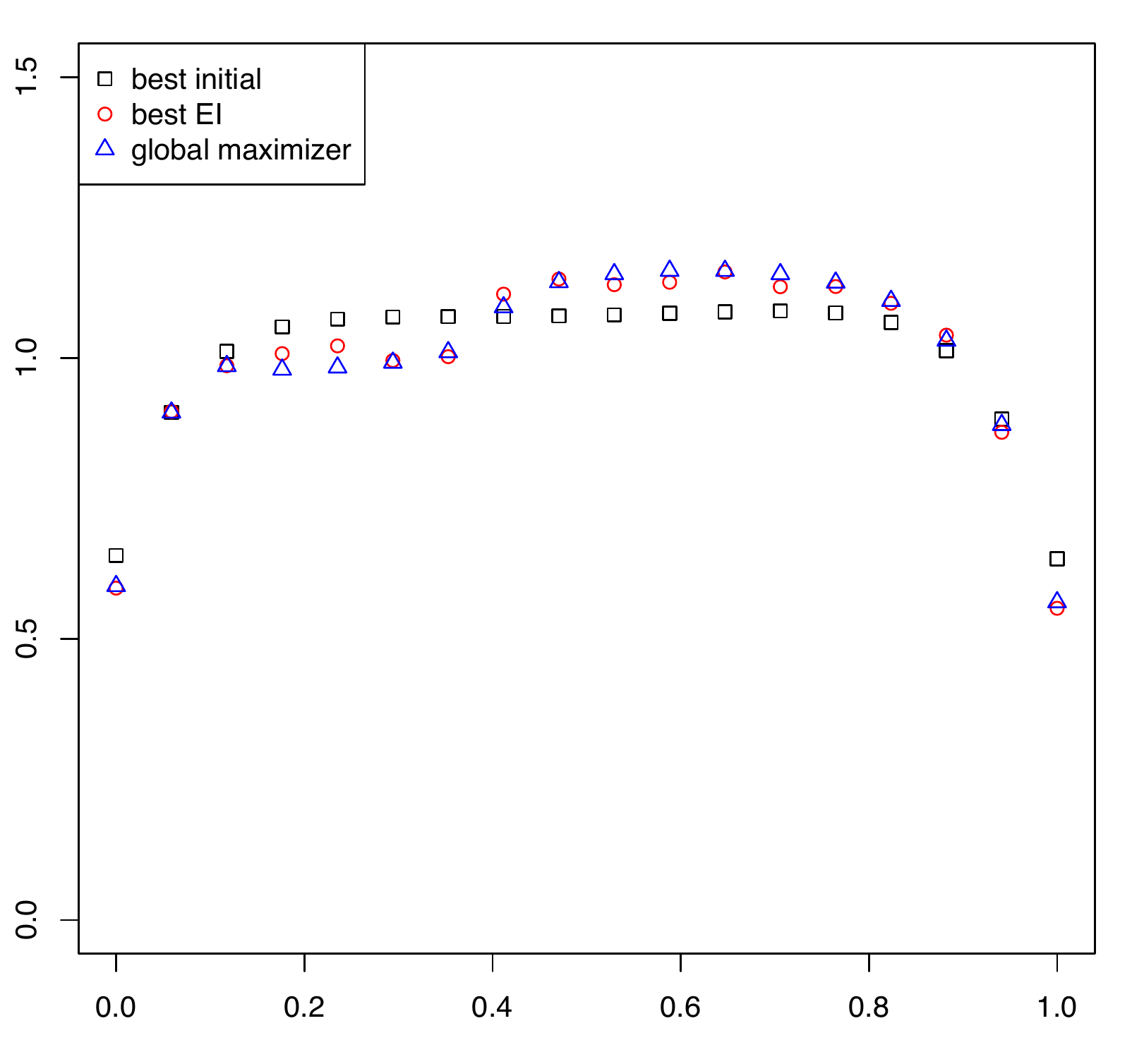}
\vspace{-.2in}
\caption{Three input curves for the 
analytic optimization problem
\eqref{eq:analytical:function}:
the true global $y_a(\bm{x})$  maximizer;
the $y_a(\bm{x})$ maximizer 
among the $100$ initial curves;  
the $y_a(\bm{x})$ maximizer 
among the $50$  curves added by 
EI/kernel density estimation.}
\label{fig:kde:toy:function:curves}
\end{center}
\end{figure}

The initial random design of $100$ curves was obtained similarly 
to that for the expert knowledge procedure above; 
the sequentially added $50$ curves 
were obtained by the EI/kernel density estimation procedure. 
The conclusions are similar to those for 
the EI/expert knowledge procedure. Namely,
among the initial $100$ training data curves, 
the maximum value of $y_a(\bm{x})$ is approximately $-0.71$
which occurs at $\bm{x}_{\mathrm{init}}$. 
With the $50$ 
additional curves obtained by expected improvement, this maximum 
increases to approximately $-0.11$
at $\bm{x}_{\mathrm{EI}}$. 
Figure \ref{fig:kde:toy:function:curves} plots the
three curves $\bm{x}_{\mathrm{init}}$,
$\bm{x}_{\mathrm{EI}}$ and $\bm{x}_0$.

\subsection{Optimizing Burn-up Credit Penalization} 
\label{subsec:optim:burnup}

In this subsection
the expected improvement procedure is carried out 
similarly as in Subsection 
\ref{section:optim:analytical} with the analytical function 
$y_a(\bm{x})$ replaced by  {\tt CRISTAL}
code function $y(\bm{x})$ evaluations. 

For both methods of identifying the valid input space, ${\cal{X}}$,
the EI algorithm was carried out starting 
from a Gaussian Process model based on $100$ observed values of 
$y(\bm{x})$.  For EI/kernel density via
the optimization problem \eqref{eq:global:optimization:kde},
these 
observed values corresponded to a subset 
$\{ \widehat{\bm{\alpha}}^{(i_1)}, \ldots , 
\widehat{\bm{\alpha}}^{(i_{100})} \}$ 
of $\{ \widehat{\bm{\alpha}}^{(1)}, \ldots , 
\widehat{\bm{\alpha}}^{(3,158)} \}$. 
This subset was selected by the following space-filling 
procedure.   First,  
$100$ barycenters where computed from a 
K means clustering algorithm applied 
to $\{ \widehat{\bm{\alpha}}^{(1)}, \ldots , 
\widehat{\bm{\alpha}}^{(3,158)} \}$.  
{Then, the $\{ \widehat{\bm{\alpha}}^{(i_1)}, \ldots , 
\widehat{\bm{\alpha}}^{(i_{100})} \}$ closest to these barycenters were selected.
The} {\tt CRISTAL}
code was run $100$ times to compute the corresponding 
$
y( \bm{x}(\widehat{\bm{\alpha}}^{(i_1)}) 
    /\bar{ \bm{x}}(\widehat{\bm{\alpha}}^{(i_1)})  )$
, \ldots,
$y( \bm{x}(\widehat{\bm{\alpha}}^{(i_{100})}) 
   /\bar{ \bm{x}}(\widehat{\bm{\alpha}}^{(i_{100})})  )$. 

For the EI/expert constraints method, a subset 
$\{ \bm{x}_{i_1}, \ldots , \bm{x}_{i_{100}} \}$ 
of $\{ \bm{x}_{1}, \ldots , \bm{x}_{3,158} \}$ 
was obtained, using the same space-filling procedure as above. The corresponding
$y(\bm{x}_{i_1}), \ldots , y(\bm{x}_{i_{100}})$  
\blue{were selected from 
the historical data base.}

The number of initial values for expected improvement, 
$100$, was 
hence selected for two reasons. First reason is that 100 initial 
observations 
allows an interpretable comparison between the results of
EI using 
expert 
knowledge versus kernel density estimation methods. The second reason was based on computational budget considerations. 
In the future, the Burn-up Credit code is expected to become more complex 
and costly to evaluate.  It was extrapolated 
that the value $100$ satisfies future budget constraints
and suggests the performance of the two input determination 
methods.  

\medskip

\noindent
{\bf EI optimization of  
Burn-up Credit Penalization}

\begin{figure}
\centering
\begin{tabular}{cc}
\includegraphics[width=7cm]{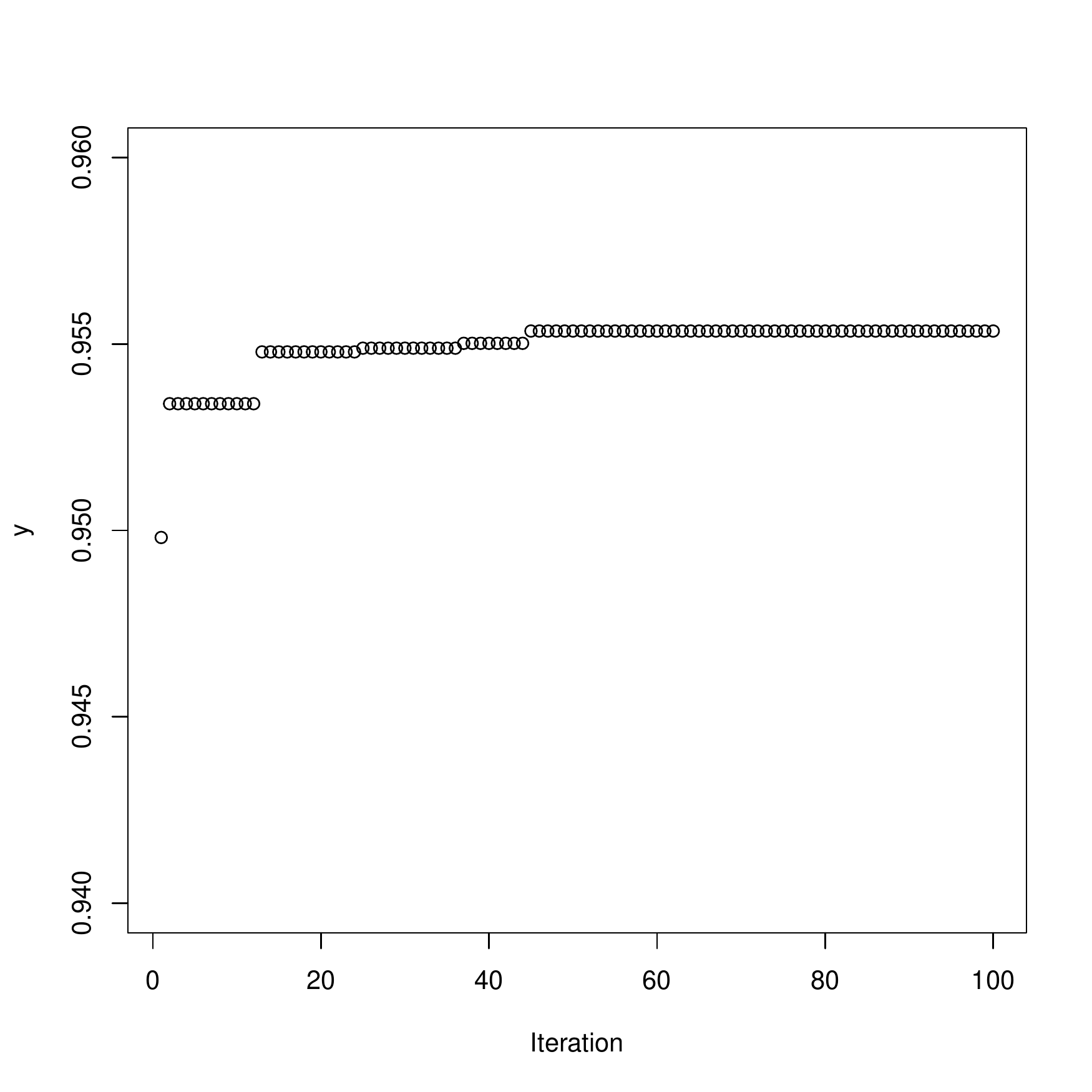}
&
\includegraphics[width=7cm]{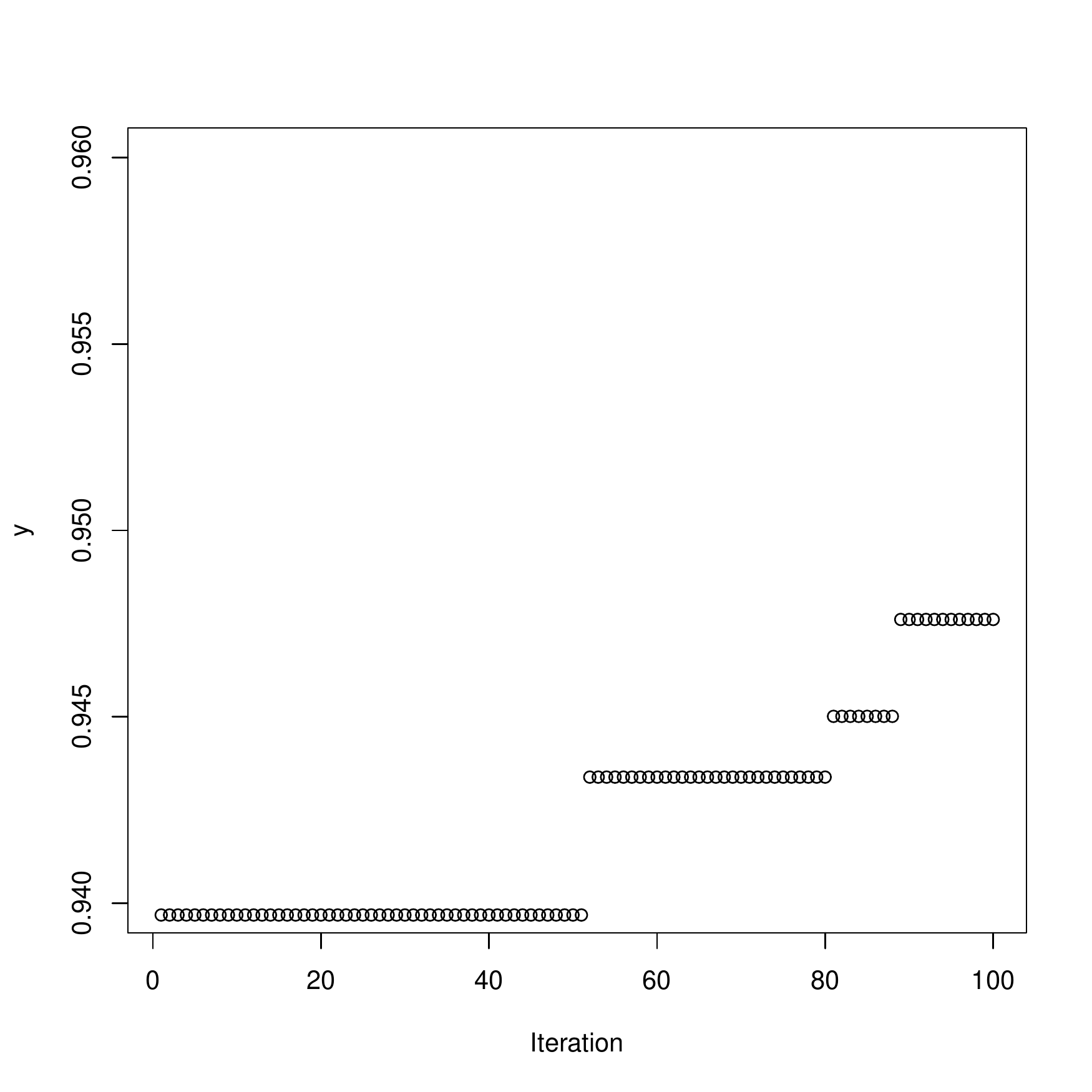}
\end{tabular}
\caption{
The cumulative maxima,  
$\max \left\{ m_{\mathrm{hist}}, y(\bm{x}_{EI,1}),\ldots,y(\bm{x}_{EI,i}) 
\right\}$, as a function of the iteration index $i$ for 
EI based on 
the expert-type constraints (left panel) and 
on kernel density estimation (right panel). 
The symbol $m_{\mathrm{hist}}$ denotes the maximum value of $y(\bm{x})$ 
among the $100$ training data input curves.}
\label{fig:cum:best}
\end{figure}

The maximum of $\{ y(\bm{x}_{1}), \ldots , y(\bm{x}_{3,158}) \}$ is equal to 
$0.94123$. This illustrates the maximization
performance using only the historical data 
base (although these values contain a small Monte Carlo noise). 
Starting with 100 training inputs and their 
corresponding $y(\bm{x})$ values and then
running $100$ 
iterations of the expected improvement procedure 
yields new curves 
$\bm{x}_{EI,1},\ldots,\bm{x}_{EI,100}$ for both the EI/expert constraints and 
EI/kernel density estimation procedures. The maximum of 
$y(\bm{x}_{EI,1}),\ldots,y(\bm{x}_{EI,100})$ is $0.95535$ 
for the EI/expert constraints procedure and is $0.94761$ for the 
EI/kernel density estimation procedure. Hence, the admissible 
set obtained from the expert knowledge is larger, so to speak, 
than that obtained from the kernel density estimation procedure, 
and allows for larger values of $y(\bm{x})$. 
This is possibly due to 
the choices of the tolerance values $\epsilon$ and of the 
distance $\Delta$ (see Section \ref{sec:computation}). 
One may also notice that, in essence, the EI/expert constraints 
procedure allows for a larger search space for optimization, 
as it does not project the curves onto a lower dimensional space.

To show the effectiveness of the two EI procedures, 
Figure \ref{fig:cum:best}
plots the cumulative maxima of $y(\bm{x})$, including that based
on the training data, for 
the $100$ EI iterations. The convergence appears to be 
relatively fast when ${\cal X}$ is determined by expert constraints. 
When ${\cal X}$ is determined by 
kernel density estimation, additional
iterations of expected improvement would likely result in a 
further improvement of $y(\bm{x})$. 

\begin{figure}[h!]
\centering
\includegraphics[width=10cm]{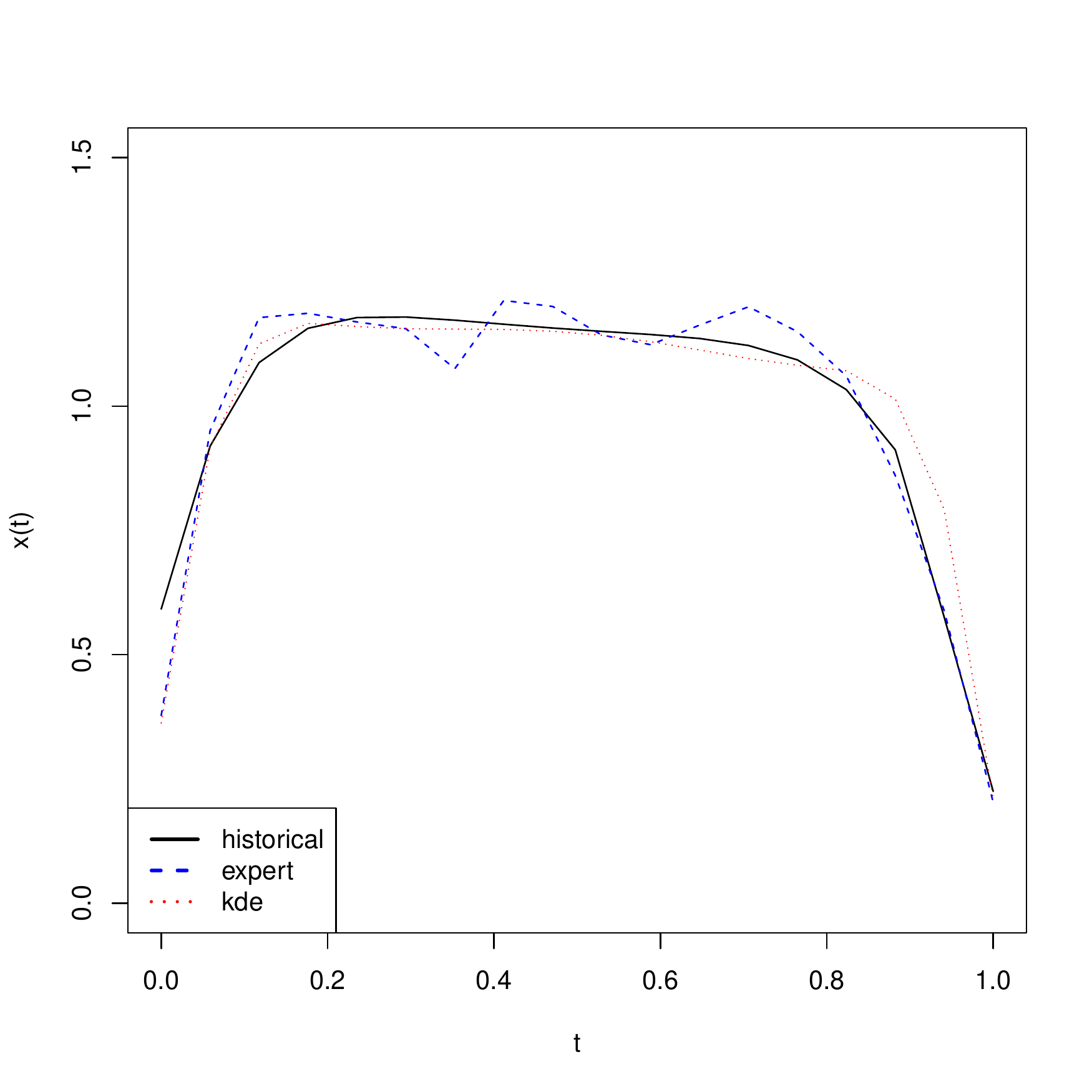}

\vspace{-.2in}
\caption{Curves that maximize the Burn-up 
Credit Penalization: (1) 
among the 3,158 historical curves;
(2) among the 100 curves added by the EI/expert constraints procedure;
(3) among the 100 curves added by the EI/kernel density estimation 
procedure.}
\label{fig:final_curves}
\end{figure}

Finally Figure \ref{fig:final_curves} plots the three curves 
$\bm{x}_{\mathrm{hist}}$, $\bm{x}_{\mathrm{expert}}$, 
and $\bm{x}_{\mathrm{kde}}$, where $\bm{x}_{\mathrm{hist}}$ 
corresponds to the maximum of the historical values 
$\{ y(\bm{x}_{1}), \ldots , y(\bm{x}_{3,158})\}$ and $\bm{x}_{\mathrm{expert}}$ (resp. $\bm{x}_{\mathrm{kde}}$) corresponds to the maximum of $y(\bm{x}_{EI,1}),\ldots,y(\bm{x}_{EI,100})$ for the EI/expert constraints (resp. EI/  kernel density estimation) procedure. The deviation from 
$\bm{x}_{\mathrm{expert}}$ and $\bm{x}_{\mathrm{kde}}$ to 
$\bm{x}_{\mathrm{hist}}$ is moderate but non-negligible. We also observe that $\bm{x}_{\mathrm{kde}}$ is smoother than $\bm{x}_{\mathrm{expert}}$, 
which is a \blue{feature}
of the spline decomposition. 

\subsection{Breath of Application of the Method and an Illustrative  Example} 
\label{subsection:general example}

Although the present methodology has been presented together with the motivating fuel rod application case, 
its level of generality goes beyond this case. Indeed, the methodology can be applied to 
any setting where the two following features are present. (1) There is an unknown function domain 
\[
\mathcal{F}
\subset 
\left\{
\bm{x} : [0,1] \to \mathbb{R}
\right\},
\]
where the domain of the functions is conventionally fixed to $[0,1]$, without loss of generality. The corresponding set of discretized curves is
\[
\mathcal{F}_d
=
\left\{
\bm{x} = (x(t_1),\ldots,x(t_d))^\top ;
x \in \mathcal{F}
\right\},
\]
for fixed grid knots $0 \leq t_1 < \dots < t_d \leq 1$.
Discretized functions $\bm{x}_1,\ldots,\bm{x}_n \in \mathcal{F}_d$ are available. (2) There is a simulator $y : \mathcal{F}_d \to \mathbb{R}$, where $y(\bm{x})$ can be evaluated (with or without noise) for any $\bm{x} \in \mathbb{R}^d$ (or any $\bm{x} $ in a fixed known subset of $\mathbb{R}^d$ containing $\mathcal{F}_d$). Each evaluation of  $y(\bm{x})$ is costly and thus the total number of evaluations is limited.

The objective is to solve the constrained optimization problem
\begin{equation} \label{eq:the:general:problem}
\max_{\bm{x} \in \mathcal{F}_d}
y( \bm{x} ),
\end{equation}
that is to simultaneously estimate the unknown
$ \mathcal{F}_d$ from the historical $\bm{x}_1,\ldots,\bm{x}_n$ and optimize $y$. The fuel rod application introduced in Section \ref{sec:MathDesc} and addressed in Subsection \ref{subsec:optim:burnup} is thus a special case of this general framework.

The methodology introduced here (that is the two methods for determining the input domain in Section \ref{sec:computation} followed by the EI procedure) can be readily applied to the general problem \eqref{eq:the:general:problem}. Note that the methodology is not restricted to the constraint of positive-valued elements in $\mathcal{F}_d$ averaging to $1$, as in the fuel rod application. Indeed, if this constraint is not present, one may just omit all the normalization steps (dividing a vector by its average).

\begin{figure}
\begin{center}
\includegraphics[width=8cm]{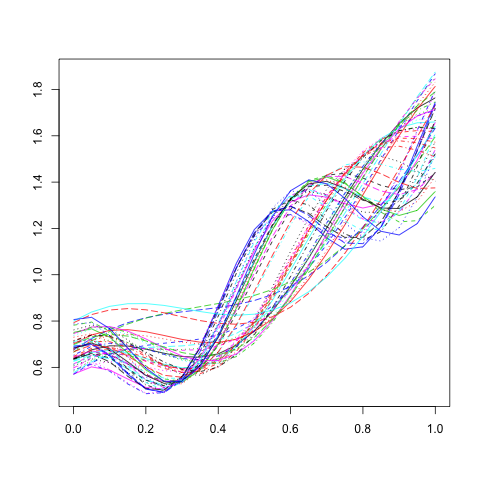}
\end{center}

\vspace{-.5in}

\caption{
Fifty historical curves for the analytical example of Subsection \ref{subsection:general example}.
}
\label{fig:general:historical:curves}
\end{figure}

Let us now provide an application of the methodology to an analytical test case.
Consider the grid knots $t_1 = 0, t_2 = 1/20 , \ldots , t_d = 1$ with $d=21$.
 Consider the function domain
\begin{equation} \label{eq:general:domain}
\mathcal{F}
=
\left\{
\bm{x}_{a,b,c}; (a,b,c) \in [4,12] \times [2,5] \times [0.8 , 1.2]
\right\},
\end{equation}
with $x_{a,b,c}(t) = (1 + \cos(at) + bt + \exp(ct)) / C_{a,b,c}$ where $C_{a,b,c} = (1/d) \sum_{i=1}^d (1 + \cos(at_i) + bt_i + \exp(ct_i))$. Hence, $\mathcal{F}_d$ is composed of vectors averaging to $1$. We consider an historical data set of size $n=1,000$ obtained by independent random sampling of elements in $\mathcal{F}_d$ by sampling $(a,b,c)$ uniformly on $[4,12] \times [2,5] \times [0.8 , 1.2]$. Figure \ref{fig:general:historical:curves} shows $50$ of these discretized functions.

We consider the curve $\bm{x}_{12,6,1}$ (slightly outside $\mathcal{F}$) and its discretized version $\bm{x}_{12,6,1}$ (slightly outside $\mathcal{F}_d$). Then the code function is defined as, for any $\bm{x} = (x_1,\ldots,x_d)^\top$,
\[
y(\bm{x})
=
- \left| \left| \bm{x} - \bm{x}_{12,6,1}  \right|  \right|_2
- 
\sin 
\left(
3
\left| \left| \bm{x} - \bm{x}_{12,6,1}  \right|  \right|_2
\right)^2.
\]
The theoretical maximizer of \eqref{eq:the:general:problem} is thus close to $\bm{x}_{12,6,1}$. The global maximum and maximizer of \eqref{eq:the:general:problem} are computed by a brute force method with $10^6$ evaluations of $y$. The maximum is approximately $-0.08$.

\begin{figure}
\begin{center}
\begin{tabular}{cc}
\includegraphics[width=8cm]{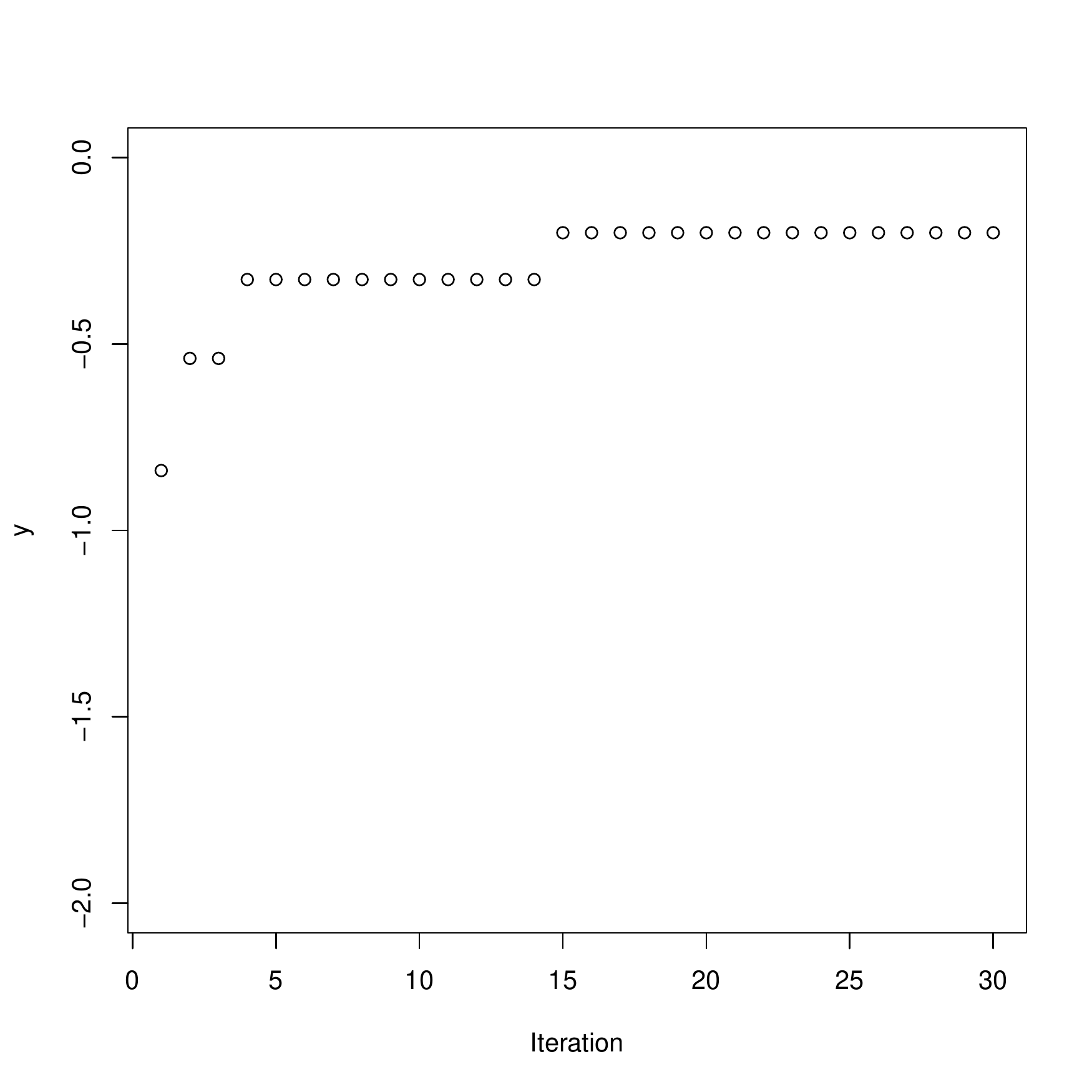}
&
\includegraphics[width=8cm]{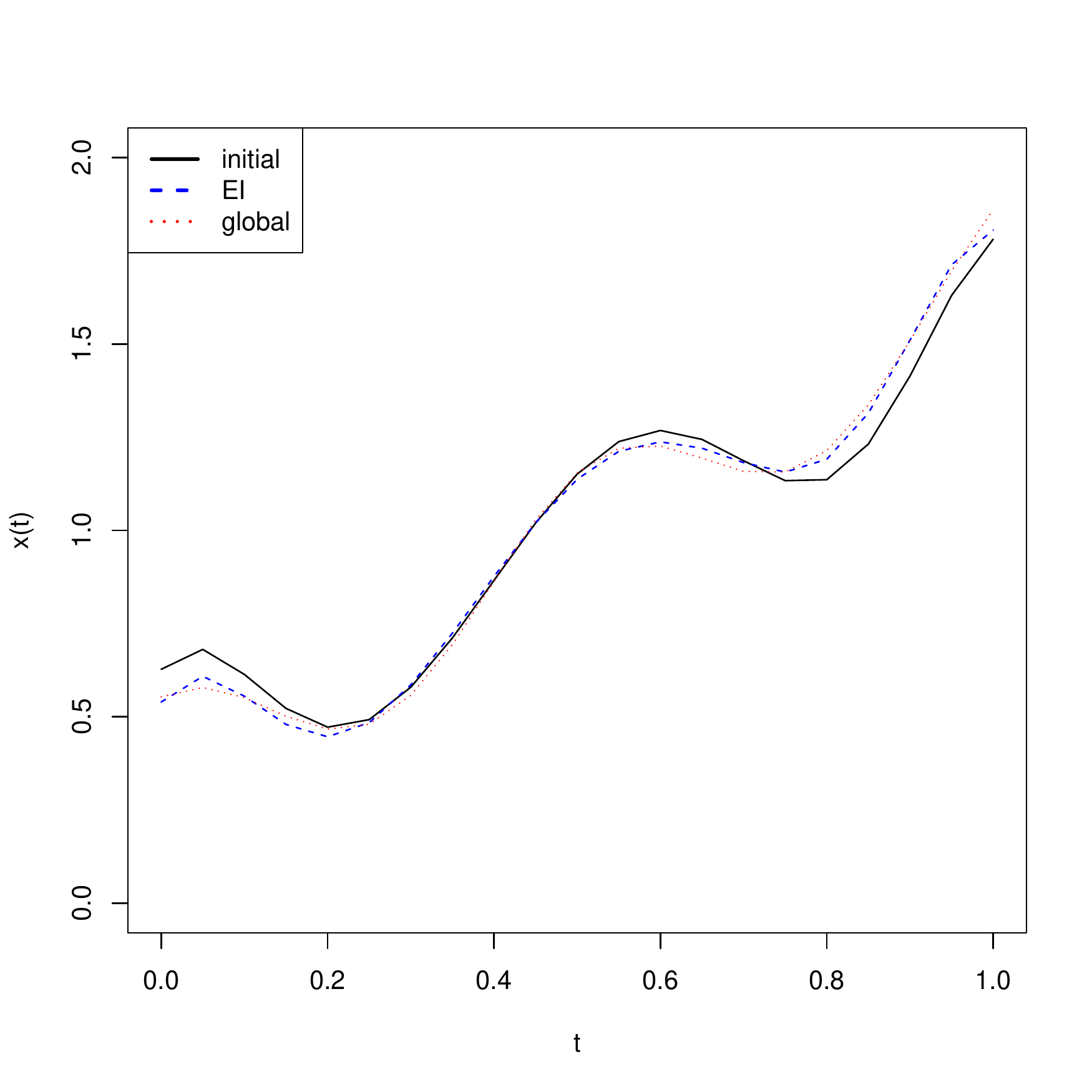}
\end{tabular}
\end{center}

\vspace{-.2in}

\caption{
Analytical example of Subsection \ref{subsection:general example} for the methodology based on projections onto a basis set followed by optimization by EI. Left: cumulative maxima of $y$ values along EI iterations. Right: best of the $30$ initial curves, the curve found by EI and the global maximizer curve.
}
\label{fig:general:kde}
\end{figure}

We first carry out the methodology based on projections onto a basis set followed by optimization by EI (Subsection \ref{subsec:kernel}).
Here there is no noise in the evaluations of $y$ nor in the Gaussian process model of EI.
Otherwise, we use the same settings as for the fuel rod application (Subsections \ref{section:determining:admissible} and \ref{subsec:optim:burnup}), in particular the same spline basis functions and the same value $\Delta = 0.05$ in \eqref{eq:4.delta}. 
For optimization, we select $30$ curves based on the same space filling construction as in Subsection \ref{subsec:optim:burnup}. We then run $30$ iterations of the EI procedure. The maximum of the values of $y$ over the $30$ initial curves is approximately $-0.84$. The maximum of the values of $y$ after the $30$ EI iterations is approximately $-0.20$. Figure \ref{fig:general:kde} shows the cumulative maxima of the values of $y$ along the EI iterations and the best of the $30$ initial curves, the curve found by EI and the global maximizer curve. The conclusion is that the methodology is successful here. With only $60$ calls to the code function, it yields a value of $y$ which is close to the maximum in \eqref{eq:the:general:problem}, and a corresponding curve which is visually very close to the maximizer in \eqref{eq:the:general:problem}. Furthermore, the curve obtained by EI is significantly closer to the global maximizer than the curve obtained by the initial space filling design. Finally, we remark that the employed methodology needs no knowledge of the nature of the set \eqref{eq:general:domain}, which is completely unrelated to the spline basis functions used.

\begin{figure}
\begin{center}
\begin{tabular}{cc}
\includegraphics[width=8cm]{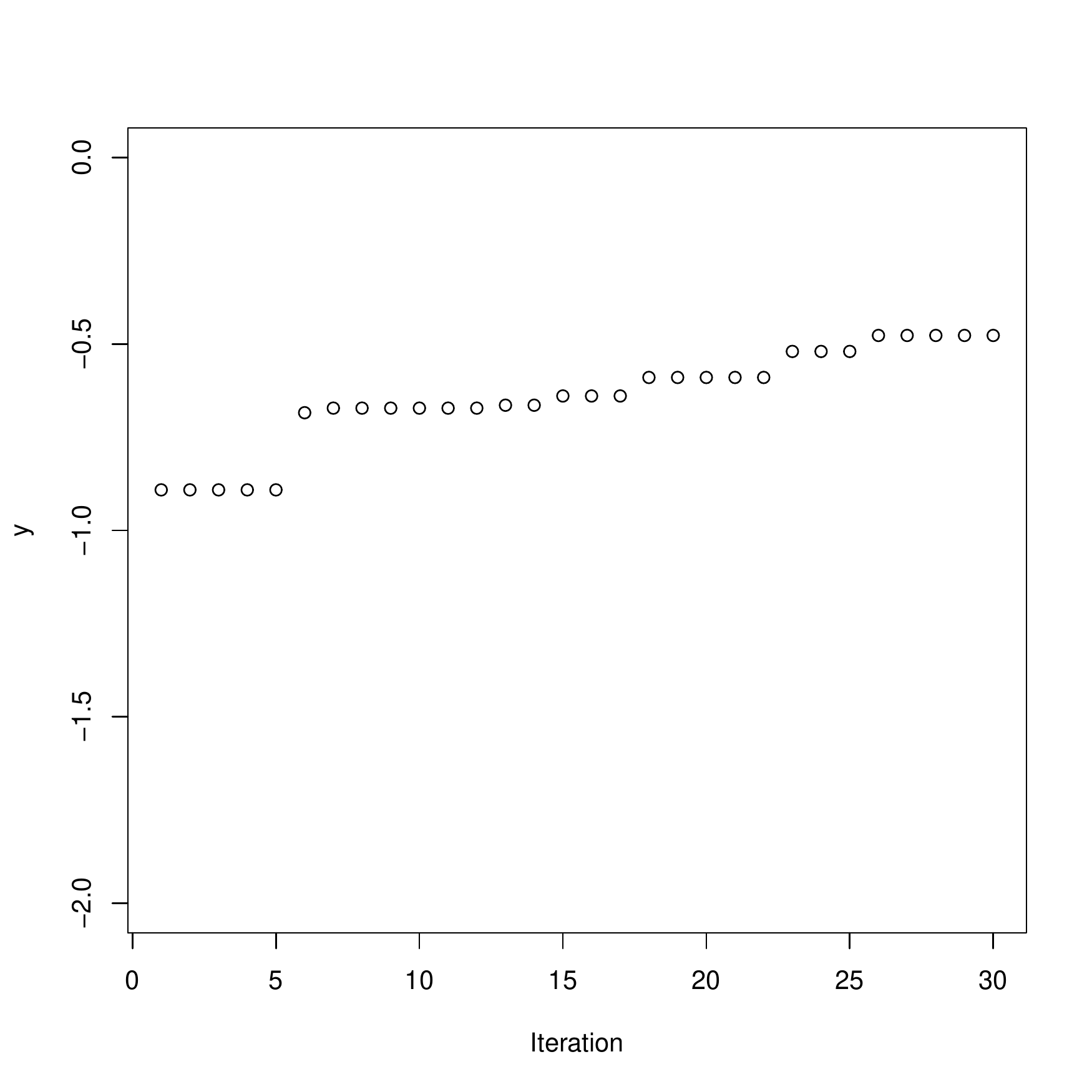}
&
\includegraphics[width=8cm]{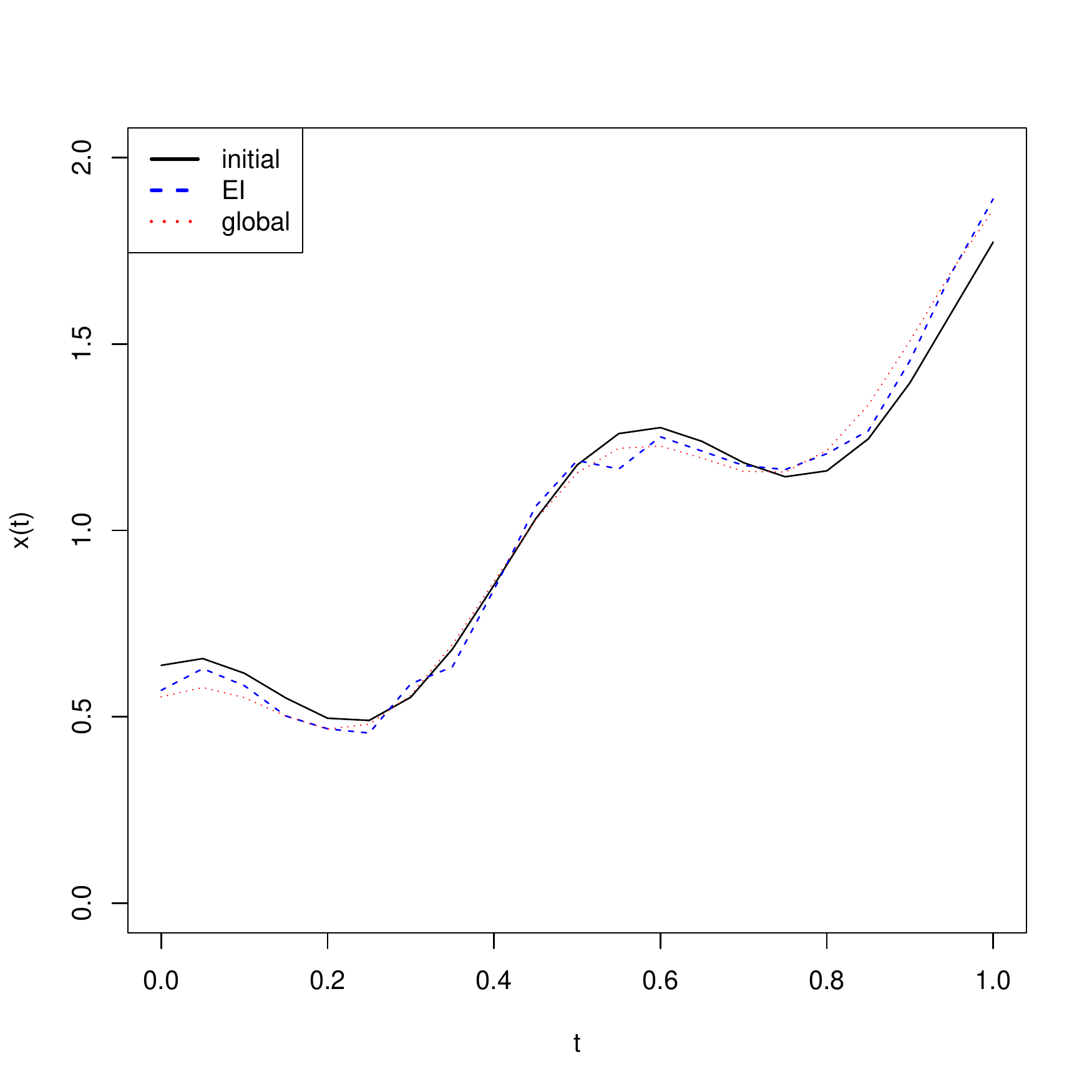}
\end{tabular}
\end{center}

\vspace{-.2in}

\caption{
Analytical example of Subsection \ref{subsection:general example} for the methodology based on expert-type constraints followed by optimization by EI. Left: cumulative maxima of $y$ values along EI iterations. Right: best of the $30$ initial curves, the curve found by EI and the global maximizer curve.
}
\label{fig:general:expert}
\end{figure}

Second, we carry out the methodology based on expert-type constraints followed by optimization by EI (Subsection \ref{subsec:expert}). 
Again, there is no noise in the evaluations of $y$ nor in the Gaussian process model of EI and, otherwise,
we use the same settings as for the fuel rod application (Subsections \ref{section:determining:admissible} and \ref{subsec:optim:burnup}), in particular the same list of constraints and tolerance levels $\epsilon$. Remark that the constraints related to the time steps $16,17,18$ for the fuel rod application naturally correspond to constraints related to the time steps $19,20,21$ here.
Again, for the optimization, we select $30$ curves based on the same space filling construction as in Subsection \ref{subsec:optim:burnup} and we run $30$ iterations of the EI procedure. The maximum of the values of $y$ over the $30$ initial curves is approximately $-0.89$. The maximum of the values of $y$ after the $30$ EI iterations is approximately $-0.48$. Figure \ref{fig:general:expert} is then similar to Figure \ref{fig:general:kde}. The conclusion that the methodology is successful also holds, similarly as previously. Compared to the methodology based on projections onto a basis set, we remark that the convergence of EI is slightly slower and that the curve found by EI is more irregular. This irregularity is similarly observed in Figure \ref{fig:final_curves} for the fuel rod application, and holds because the expert knowledge methodology does not project the curves onto function spaces, but instead treats them as $d$-dimensional vectors.

In summary, Figures \ref{fig:general:kde} and \ref{fig:general:expert} illustrate the robustness of the present suggested methodology. Indeed, the methodology, as calibrated for the fuel rod application (in particular the choice of the spline projection functions, of the expert-type constraints and of various numerical parameters), proves to be efficient in the new setting \eqref{eq:the:general:problem}. Of course, in other specific settings, other choices of calibration (for instance other expert-type constraints) may lead to even better results.

Let us conclude Subsection \ref{subsection:general example}, by further discussing the generality of the present framework and methodology. In \eqref{eq:the:general:problem}, we have presented a general framework where optimization over an unknown domain of one-dimensional curves is considered. In the fuel rod application, these one-dimensional curves correspond to spatial dependence, since they provide the spatial distribution of the burn-up rate. 

In \eqref{eq:the:general:problem}, the $d$ grid knots are the same for all the curves, which is the case for the fuel rod application motivating the present work. While the case of identical grid knots among the historical curves is relatively common, it may happen that each curve is discretized on a different set of knots. In this case, as mentioned in Section \ref{sec:MathDesc}, various interpolation schemes can be applied. If each curve is discretized at a large number of knots, then function approximation principles suggest that the impact of the interpolation scheme will be small. If some curves are discretized at a small number of knots, then the choice of the interpolation scheme may influence the final results. Studying this influence in further real examples deserves to be the topic of future work.

The framework and methodology could be extended to optimizing over an unknown domain of two-dimensional surfaces (typically if a temporal dependence is also tackled) or higher dimensional functions. Indeed, one may still define constraints as in Subsection \ref{subsec:expert}, for instance using finite differences in various directions, or extending the notion of total variation to the multi-dimensional case.
Similarly, as in Subsection \ref{subsec:kernel}, one may decompose the surfaces or functions on finite-dimensional bases of multi-dimensional functions.

When considering an unknown domain of multi-dimensional surfaces, the optimization space becomes more complex, and applying EI can become more challenging. Specific procedures for high-dimensional optimization based on EI, for instance \cite{Salem2019}, may be relevant in this case. Depending on the physical application, other aspects may be present and yield further complexity and multi-dimensional heterogeneity.
While the current results indicate that our suggested methodology can be applicable in the various more complex settings discussed above (as it is robust as discussed above), in future work, it would be valuable to further test it in these settings.

\section{Summary and Discussion} 
\label{SummaryDiscussion} 

This paper proposes methods to identify a potentially
complicated input optimization domain  which is known to be a subset of the simplex, based
on observational historical data that are
known to belong to the input domain.  
It also shows how a variant of the
EGO algorithm for deterministic output 
can be applied to optimize
the mean output of a stochastic simulator over this 
domain.  The 
expected improvement function is maximized over 
an input region of positive Lebesgue measure by applying
a linear
transformation of the simplex to a \blue{lower-dimensional}
space. 

The application of these methods \blue{ to a large}
validated historical database of burn-up 
profiles is \blue{an original} proposal to 
solve the problem of burn-up credit in nuclear safety \blue {assessment}. 
It has to be compared with \blue{current}
approaches, \blue{such as ones that use}
pre-defined profiles to check the sub-criticality of burned assemblies. 
More broadly, in applications where it tends to be 
difficult to pre-define reference profiles (like in mixed oxides fuels), 
a more general approach like the one presented here should 
be more robust.

We conclude by mentioning  two \red{problems} that have not been addressed in this paper
but are topics for future research.  \red{The first problem stems from the} 
\blue{frequently-occurring} need 
in climate science and other scientific areas to 
 build adaptively an input domain from training data.
Climate models consist of submodels for surface temperatures,
wind magnitude, wind  velocity, sea surface temperatures
and other interacting constituents that
determine the climate. These submodels
must all be computable and verifiable.
The  bounds on the input domain where   
all the composite models can be
simultaneously run is {\em unknown} and can be complex.  
Thus the problem of identifying the
input region is one of sequential
design.  A series of
 inputs is identified with the resulting attempted model 
run being successful or not.  These 
data are used to estimate the input domain. 
The \red{second problem} is the determination of
sensitivity analysis tools for the mean of 
a stochastic simulator when the 
input domain
is an estimated subset of the simplex.  
\blue{The research of  \citet{LoeWilMoo2013} who developed 
global sensitivity tools for deterministic simulator output defined on the simplex
is a starting point for this more complicated scenario.}


\vspace{-.4in}
 \begin{center}
 \item \section*{ACKNOWLEDGMENTS}
 \end{center}

 \vspace{-.1in}

 \doublespacing

The authors wish to thank two anonymous referees, for suggestions which led to an improvement of this paper.

This research was conducted with the support of the Chair in Applied Mathematics
OQUAIDO, gathering partners in technological research (BRGM, CEA, IFPEN, IRSN,
Safran, Storengy) and academia (CNRS, Ecole Centrale de Lyon, Mines \blue{Saint-\'{E}tienne}, 
University of Grenoble, University of Nice, University of Toulouse) in the development of
 advanced methods for Computer Experiments.

The authors would like also to thank the Isaac Newton Institute for
Mathematical Sciences for support and hospitality during the programme on
{\em Uncertainty Quantification} when work on this paper was undertaken and
 The Statistical and Applied Mathematical Sciences Institute
for support and hospitality during the program
{\em Model Uncertainty: Mathematical and Statistical}.
Finally, this work was supported by:
EPSRC grant numbers EP/K032208/1 and EP/R014604/1.
This research was also
sponsored, in part, by the National Science Foundation under Agreements  DMS-0806134 and DMS-1310294 (The Ohio State University).
Any opinions, findings, and conclusions or recommendations expressed in this material are those of the authors and do not
necessarily reflect the views of the National Science Foundation.


\end{document}